# Multiple fluid theory of cosmic evolution and its thermodynamic analysis

**Shouvik Sadhukhan[1] Alokananda Kar[2] Surajit Chattopadhyay[3]**

1. Indian Institute of Technology; Kharagpur-721302; Department of Physics; West Bengal; India
    Email: shouvikphysics1996@gmail.com
2. University of Calcutta; Department of Physics; Kolkata-700009; West Bengal; India
    Email: alokanandakar@gmail.com
3. Department of Mathematics, Amity University Major Arterial Road, Action Area II, Rajarhat New Town, Kolkata 700135, India
    Email: surajitchatto@outlook.com

**Abstract**

In this paper we have discussed the modified gravity and scalar field DE model specifically DBI essence model with the analysis of thermodynamics during cosmological evolution. We have used the modified gravity with the form $f(R,T) = R + 2\lambda T$ in our calculations. The basic aim behind this paper is to discuss a theory that unifies modified gravity with DE models including the solutions of some cosmological problems like thermodynamics energy conditions violation problems, finite time future singularity problems, initial singularity problem, Cosmic inflation problem, decelerated expansion problem, graceful exit problem, reheating problem, bouncing nature problem, phase transformation-spontaneous symmetry breaking problem, negative heat capacity paradox problem and obviously the present day universe problems (Continuously decreasing temperature problem, present day DE dominated expansion problem). We have established the viscous effects (both positive and negative viscosity) in our calculation and discussed the negative-positive viscosity by introducing two special type of energy cycles. Finally, we have discussed the stability conditions for universe evolution through cosmic perturbation and resolved the instability problems. We have also shown the state finder trajectories for both with and without viscous fluid on the basis of our calculations to compare our research with the results of other independent DE models.

## 1. Introduction

In the context of cosmic evolution, the acceleration is one of the biggest question. Different models have been studied to establish the cosmic acceleration to resolve the big three problems i.e. the horizon problem, magnetic monopole problem and flatness problem. Dark energy and the modified gravity are the one to establish the both inflationary and late time acceleration. [1,2]

Modified gravity is mostly used as a substitute of dark energy and exotic matters with the introduction of higher order Ricci scalar modifications. Different types of modified gravity models are available in literature. The higher order Ricci scalar modifications are generally called as local modifications as they follow local gauge transformations. On the other-hand few kinds of non-local modifications are also interpreted in literature to merge the ideas of string theory and gravity geometry. In the list of local modifications Gauss Bonnet, $f(R,T)$ gravity is mostly included. The $f(T)$ gravity and $f(Q)$ gravities are also included in teleparallel and non-metric gravities respectively. The accelerating universe and cosmic stability can be clearly established with such kind of models. [3-12]

In general, dark energy can be discussed in two different ways. They are static and Dynamic dark energy models. Positive cosmological constant model is the earliest static dark energy model. But static model failed to explain the present time vacuum energy ambiguity (cosmic coincidence problem). That's why the idea of dynamical dark energy models was introduced. Till date three types of dynamic dark energy models are available in literature, namely scalar field model (Quintessence, DBI-Essence, Tachyon etc.), Fluid model (Chaplygin gas) and Holographic model. Here in this paper we'll be specifically discussing DBI-Essence dark energy model, which is one of the scalar field model where the brane tension has been introduced. The cosmic inflationary phase transition is well explained with tachyonic dark energy model but it has failed to discuss the brane bulk tension and DBI-essence scalar field generalized this problem. Since, both modified gravity and dark energy can explain accelerated expansion of universe we will try to establish a correspondence relation between the geometry and dark energy. [19-28, 29-36, 38-42]

After including the modifications in Ricci tensor of Einstein action, one can attempt to unify all the cosmic phases with the single unified action. Several literatures are available in this context. [3-12] Different cosmic phases are controlled by different modifications of the Einstein tensors. All the phases can be added to establish a unified action. However, it is difficult to establish the phase transitions between different phases with this unified action. In this work we have attempted to introduce a technique that can unify all the phases as well as provide conditions of phase transitions. The necessary decelerating phase after the cosmic inflation can also be obtained by our technique.

To check the thermodynamics stability of a model we must satisfy the energy condition and calculate the energy parameters. The energy conditions of dynamical fluid is obtained from the famous Raychaudhuri equation. According to the equation the accelerating universe is only possible when strong energy condition is violated. [43-55] The stability parameters are obtained from the perturbation stability analysis. [56-61] A cosmological model should be stable with respect to geometric perturbation. In thermodynamics the change in entropy and heat capacity should be positive for an accelerating universe. Therefore, thermodynamics plays a major role in the stability of our model. Negative heat capacity paradox is one of the unsolved problem in thermodynamics. In our work we have addressed this unsolved problem of negative heat capacity.

our unification scheme is based on Multiple fluid cosmology. Generally, in this model the evolution of universe is explained by considering radiation, dust, exotic matters, dark energy etc. as the components. [37] However in this paper we explore the multiple fluid cosmology in terms of different fluid model available in the literature, such as inhomogeneous fluid model, viscous fluid model, modified gravity geometry and scalar field models. We have reconstructed the scale factor which can explain all the phases of evolution. [13-18]

Cosmic interaction is one of the most interacting and successful process to establish the phase transitions. During the cosmic phase transitions there should have some major changes in the energy picture of the fluid systems. Hence this change can only be possible with interpretation of the fluid interaction. On the other hand, in compact bodies the interaction can provide additional pressures to establish equilibrium. [62-63, 64-80]

This paper has been prepared as follows.

In section 2 and 3 we have discussed the basics of modified gravity and DBI-Essence scalar field models. In section 4, 5 and 6 we discussed the basics of singularity problems and thermodynamics energy conditions. In section 7 the thermodynamics for f(R,T) with and without inhomogeneity has been discussed. In section 8 and 9 we established the multiple fluid mechanism and thermodynamic cycles to investigate the cosmic phases. Section 10 and 11 we discussed the stability analysis with cosmic perturbation

## 2. General basic calculations on modified gravity

The generalized action for the modified gravity is [74, 75]

$$S = \frac{1}{16\pi} \int f(R, T)\sqrt{-g}d^4x + \int L_m\sqrt{-g}d^4x \qquad (1)$$

Now using the least action principle w.r.t variable as metric tensor we get the field equation as,

$$f_R(R,T)R_{\mu\nu} - \frac{1}{2}f(R,T)g_{\mu\nu} + (g_{\mu\nu}\Box - \nabla_\mu\nabla_\nu)f_R(R,T) = 8\pi T_{\mu\nu} - f_T(R,T)T_{\mu\nu} - f_T(R,T)\Theta_{\mu\nu} \qquad (2)$$

Now from further calculation we get $\Theta_{\mu\nu} = -2T_{\mu\nu} - pg_{\mu\nu}$ so the final form of Einstein's modified field equation will be

$$f_R(R,T)R_{\mu\nu} - \frac{1}{2}f(R,T)g_{\mu\nu} + (g_{\mu\nu}\Box - \nabla_\mu\nabla_\nu)f_R(R,T) = 8\pi T_{\mu\nu} - f_T(R,T)T_{\mu\nu} + f_T(R,T)[2T_{\mu\nu} + pg_{\mu\nu}]$$

(3)

Now the equation (3) can be re-write for generalized modified action as follows. (using FRW model)

$$\rho + 3H(\rho + p) f_T(R, T) + \tfrac{1}{2} f(R, T) + 3(\dot{H} + H^2) f_R(R, T) - 3H(\dot{R} f_{RR}(R, T) + g_{\mu\nu} \dot{T} f_{RT}(R, T)) = 0 \quad (4)$$

From [68-69] we'll discuss modified structure of $f(R, T) = R + 2f(T) = R + 2\lambda T$. Assuming $f(T) = \lambda T$.

We'll use FRW curvature free line element as

$$ds^2 = dt^2 - a^2(t)(dx^2 + dy^2 + dz^2) ; \quad (5)$$

The Energy momentum tensor as $T_{\mu\nu} = (\rho_m + p_m) u_\mu u_\nu - p_m g_{\mu\nu}$

We have also considered $8\pi G = 1$

So, the Friedmann equation corresponding to our choice of modified gravity model are as follows,

$$3H^2 = \rho_m + 4(\rho_m + p_m)\lambda + 2\lambda T \quad (6a)$$

$$2\dot{H} + 3H^2 = -p_m + 2\lambda T \quad (6b)$$

Where $\rho_f = 4(\rho_m + p_m)\lambda + 2\lambda T$ and $p_f = -2\lambda T$

So $w_f = -\dfrac{2\lambda T}{4(\rho_m + p_m)\lambda + 2\lambda T + \rho_m}$ which will provide us the informations about the phase transformations in cosmic evolution.

So, we have derived the pressure, density and also the EOS parameter for our choice of modification. Now these results will be used to discuss the correspondence between this modification and DBI scalar field.

Here we'll consider the following forms of the parameters necessary for our calculation. Here we considered the universe to follow power law expansion as used in the paper reference [38]. To remove the initial singularity, we replaced $a \propto t^m$ with $a \propto (a1 + bt)^m$. Where $m > 1$ for accelerated expansion.

$R = 6(2H^2 + \dot{H})$ ; $T = \rho_m - 3p_m$ ; $p_m = w_m \rho_m$ ; $\rho_m = \rho_{m0} a^{-3(1+w_m)}$ ; $a = a_0(a1 + bt)^m$

Here $R$ is the Ricci scalar, $T$ is the trace energy tensor and $\rho_m$ is the dark matter energy density.

Here we have decomposed the effective energy momentum tensor into dark matter and modified gravity components. So, the effective energy momentum tensor part can be written as follows. [13, 27-34]

$$T_{\mu\nu}^{eff} = (\rho_m + \rho_f + p_m + p_f) u_\mu u_\nu - g_{\mu\nu}(p_m + p_f)$$

(7a)

If the modified gravity is further modified with viscosity, we may write as follows.

$$T_{\mu\nu}^{eff} = (\rho_m + \rho_f + p_m + p_f - 3\eta H)u_\mu u_\nu - g_{\mu\nu}(p_m + p_f - 3\eta H)$$

(7b)

So, we get the following calculations.

### 2.1. Without viscosity:

The energy density can be written as;

$\rho_f = 4(\rho_m + p_m)\lambda + 2\lambda T$

Or, $\quad \rho_f = 4(1 + w_m)\lambda \rho_{m0} a_0^{-3(1+w_m)} (a1 + bt)^{-3(1+w_m)m} + 2(1 - 3w_m)\lambda \rho_{m0} a_0^{-3(1+w_m)} (a1 + bt)^{-3(1+w_m)m}$ (8)

and the pressure can be written as;

$p_f = -2\lambda T$

Or, $p_f = -2(1 - 3w_m)\lambda \rho_{m0} a_0^{-3(1+w_m)} (a1 + bt)^{-3(1+w_m)m}$ (9)

So, we get;

$$w_{tot} = \frac{-2(1-3w_m)\lambda\rho_{m0}a_0^{-3(1+w_m)}(a1+bt)^{-3(1+w_m)m}}{4(1+w_m)\lambda\rho_{m0}a_0^{-3(1+w_m)}(a1+bt)^{-3(1+w_m)m}+2(1-3w_m)\lambda\rho_{m0}a_0^{-3(1+w_m)}(a1+bt)^{-3(1+w_m)m}+\rho_{m0}a_0^{-3(1+w_m)}(a1+bt)^{-3(1+w_m)m}}$$ (10)

From the above expression we observe $w_{tot}$ gives the constant value with time. In order to explain all the phase transition in the evolution process we must have time varying $w_{tot}$. Therefore, we introduced the idea of viscosity to investigate the nature of $w_{tot}$.

### 2.2. With viscosity:

Here if we introduce the viscosity in the system, we may get the modifications in the variables as follows.

For coefficient of viscosity $\eta$ we may write as follows. [13, 27-34]

$$3\eta H = \eta_0(t)(3H)^{n+1} = 3^{\frac{n+1}{2}}\eta_0(t)(3H^2)^{\frac{n+1}{2}} = 3^{\frac{n+1}{2}}\eta_0(t)(\rho_m + \rho_f)^{\frac{n+1}{2}}.$$

So, we get,

$p'_f = p_f - 3\eta H = p' - 3^{\frac{n+1}{2}}\eta_0(t)(\rho_m + \rho_f)^{\frac{n+1}{2}}$

Or, $p'_f = -2(1 - 3w_m)\lambda \rho_{m0} V^{-(1+w_m)} - 3^s \eta_0(t)(\rho_{m0} + (6 - 2w_m)\lambda \rho_{m0})^s V^{-s(1+w_m)}$ (11)

Density will be,

$$\rho_f = (6 - 2w_m)\lambda\rho_{m0}V^{-(1+w_m)} \tag{12}$$

And EOS parameter will be as follows.

$$w_{tot} = -\frac{2\lambda}{6\lambda+1} - 3^s\eta_0(t)\rho_{m0}^{s-1}(1+6\lambda)^{s-1}V^{-(s-1)} \tag{13}$$

So, the graphs will be as follows for $w_m = 0$. (Fig. 1 to 9)

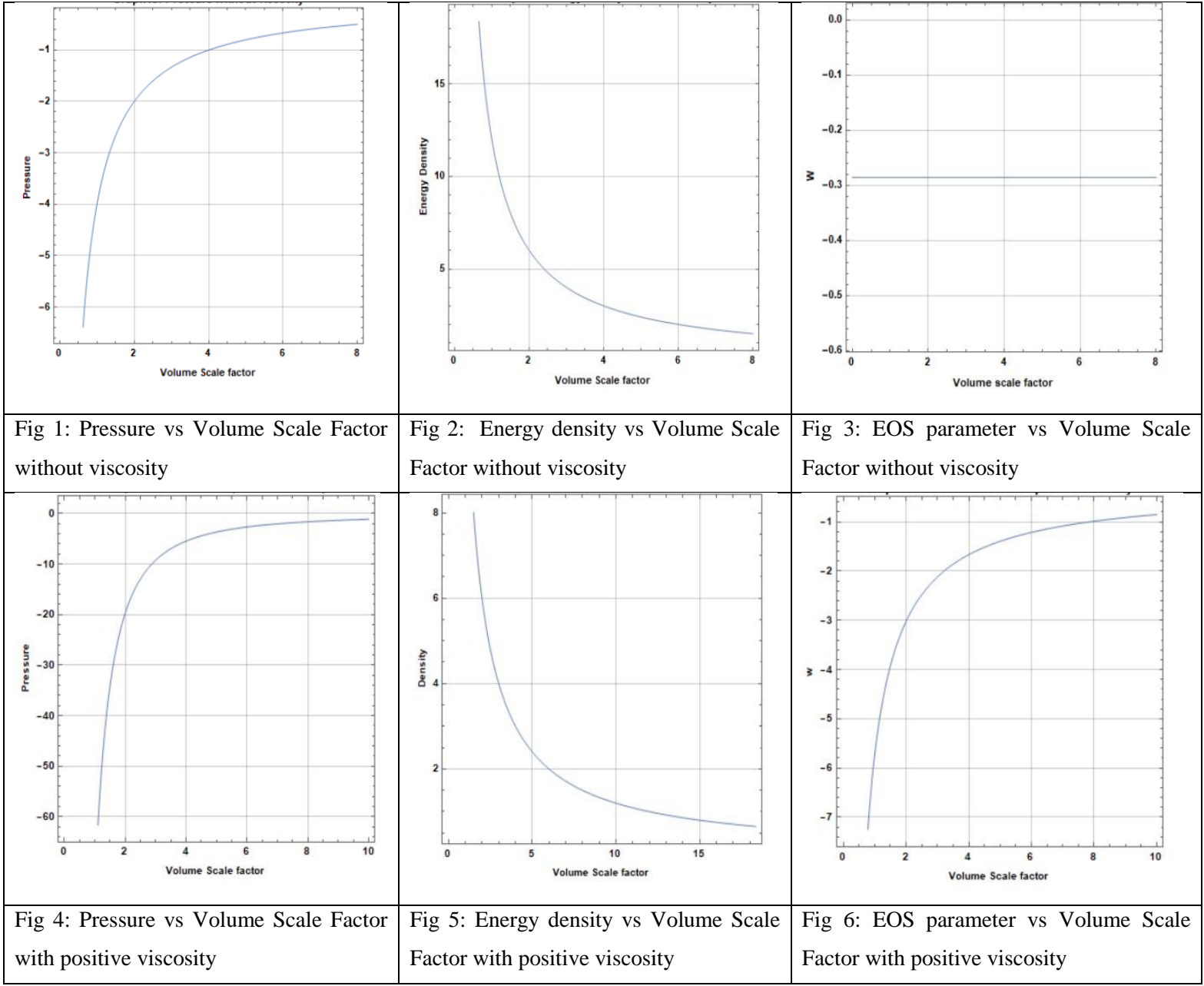

| Fig 1: Pressure vs Volume Scale Factor without viscosity | Fig 2: Energy density vs Volume Scale Factor without viscosity | Fig 3: EOS parameter vs Volume Scale Factor without viscosity |

| Fig 4: Pressure vs Volume Scale Factor with positive viscosity | Fig 5: Energy density vs Volume Scale Factor with positive viscosity | Fig 6: EOS parameter vs Volume Scale Factor with positive viscosity |

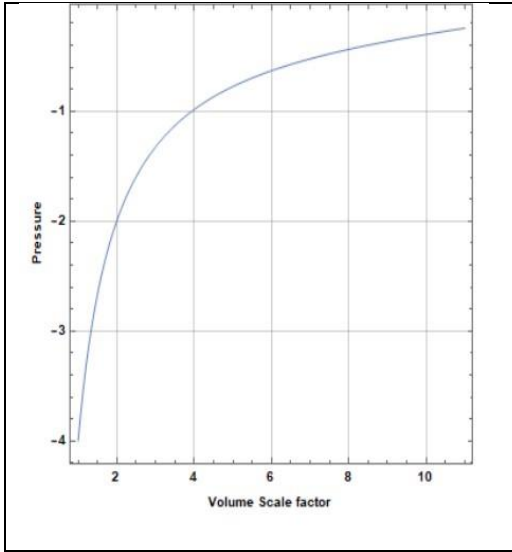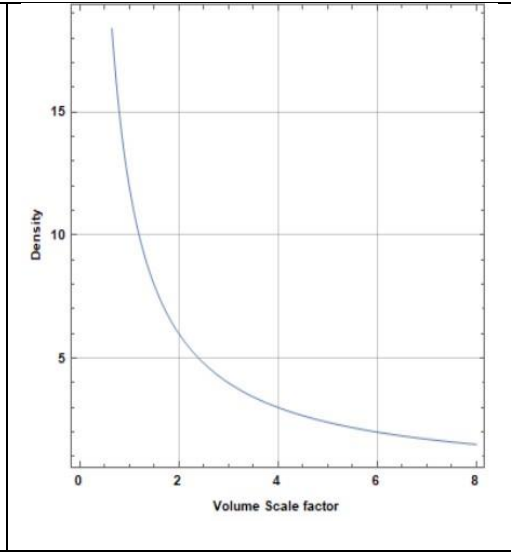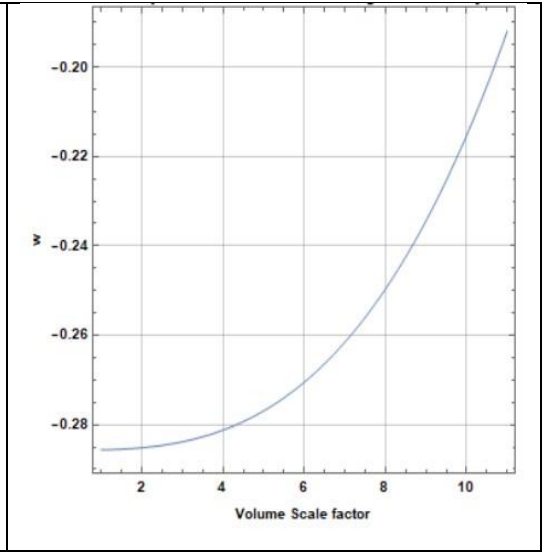

| Fig 7: Pressure vs Volume Scale Factor with negative viscosity | Fig 8: Energy density vs Volume Scale Factor with negative viscosity | Fig 9: EOS parameter vs Volume Scale Factor with negative viscosity |

In figure 3 we observe constant EOS parameter for perfect fluid system. In figure 6 and 9 we get time varying EOS parameter with the introduction of viscosity. Here we have introduced both positive and negative viscosity ($\eta < 0$ for negative viscosity and $\eta > 0$ for positive viscosity). The advantages of negative viscosity will be discussed in later section.

### 3. DBI Essence Scalar Field Dark energy model:

Here we will discuss the DBI-Essence scalar field dark energy model. This is a generalized scalar field model which will introduce Brane tension term in the tachyonic model to give Bulk-Brane cosmology. We'll use this model to find a scalar field corresponding to $f(R,T)$ modified gravity. We'll also use the corresponding kinetic energy and potential energy in our later sections to discuss the thermodynamics. We have assumed the potential that follows the power law rule.

Now the DBI Essence action can be written as [54]

$$s_{dbi} = \int d^4x \, a^3(t)[T(\varphi)\sqrt{1 - \frac{\dot\varphi^2}{T(\varphi)}} + V(\varphi) - T(\varphi)] \tag{14}$$

So, we get

$$\rho_{dbi} = (\gamma - 1)T(\varphi) + V(\varphi) \tag{15a}$$

$$p_{dbi} = \left(1 - \frac{1}{\gamma}\right)T(\varphi) - V(\varphi) \tag{15b}$$

Where $\gamma = [1 - \frac{\dot{\varphi}^2}{T(\varphi)}]^{-\frac{1}{2}}$

Here we'll consider generalized condition on the choice of $\gamma$ and $T(\varphi)$. Suppose $V(\phi) = T(\varphi) = m_1 \dot{\varphi}^n$. From above calculation we get $T(\varphi) = \frac{\gamma}{\gamma^2 - 1}[\rho_{dbi} + p_{dbi}]$

So, using $T(\varphi)$ we get the following differential equation.

$$m_1 \dot{\varphi}^4 + [\rho_{dbi} + p_{dbi}]^2 [\dot{\varphi}^{2-n} - m_1] = 0 \tag{16a}$$

Now substituting $\rho_{dbi}$ and $p_{dbi}$ using $\rho_f$ and $p_f$ from above calculations we can get the corresponding scalar field of modified gravity. So, the final differential equation will be as follows.

$$m_1 \dot{\varphi}^4 + [\rho_f + p_f]^2 [\dot{\varphi}^{2-n} - m_1] = 0 \tag{16b}$$

Now we'll derive the above equations and plot the graphs for scalar field, scalar field potential and EOS parameter.

Now consider n = 2 and we get as follows.

The kinetic term is;

$$\frac{1}{2}\dot{\phi}^2 = \frac{1}{2}(\rho_f + p_f)\left(1 - \frac{1}{m_1}\right)^{\frac{1}{2}} \tag{16c}$$

And the potential energy will be;

$$V(\phi) = \frac{1}{2}(\rho_f - p_f) + (1 - 2m_1)(\rho_f + p_f) \tag{16d}$$

From the definitions of pressure and density in equations (8-13) the graphs can be drawn as follows, using $w_m = 0$. For both with and without viscosity case. (Fig. 10 to 15)

| | |
|---|---|
| 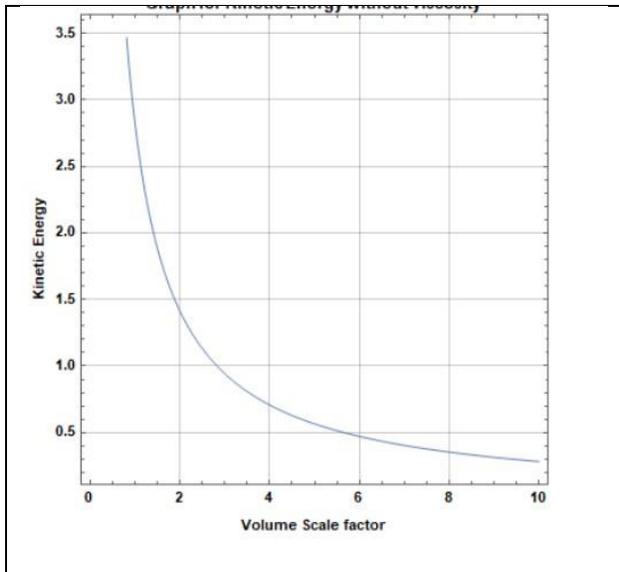 Fig 10: Kinetic Energy vs Volume scale factor Without Viscosity | 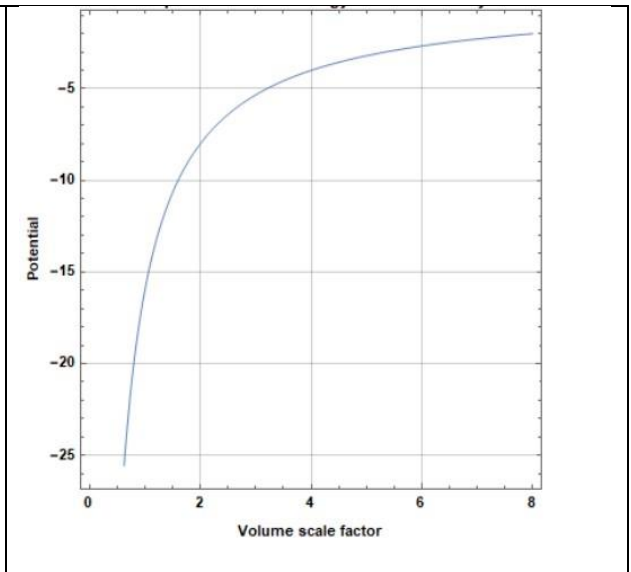 Fig 11: Potential Energy vs Volume Scale Factor Without Viscosity |
| 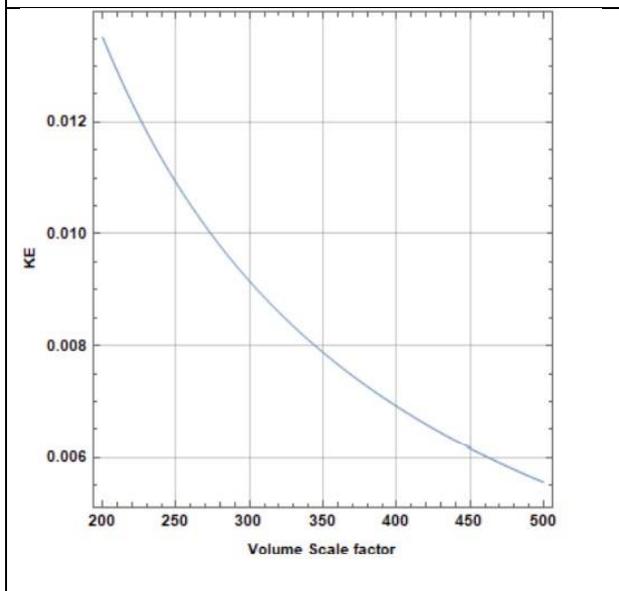 Fig12 : Kinetic Energy vs Volume scale factor With positive Viscosity | 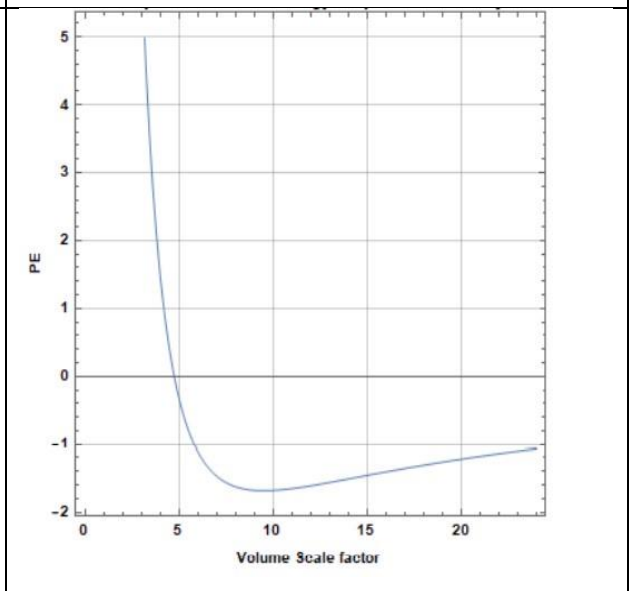 Fig 13 : Potential Energy vs Volume Scale Factor With positive Viscosity |

| 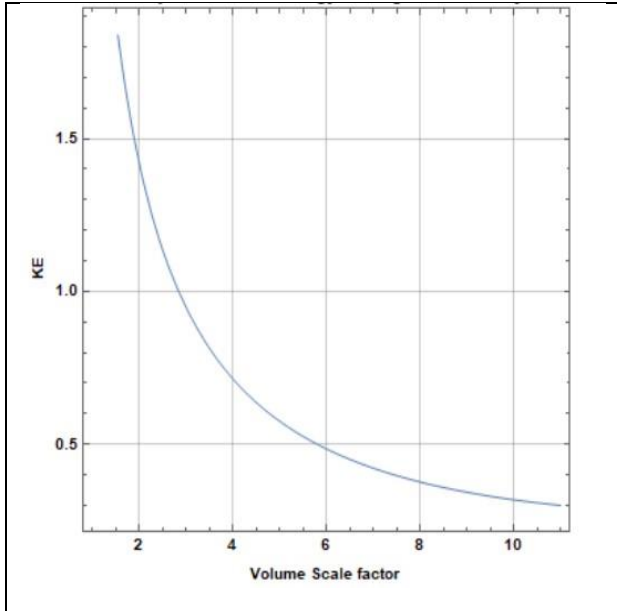 | 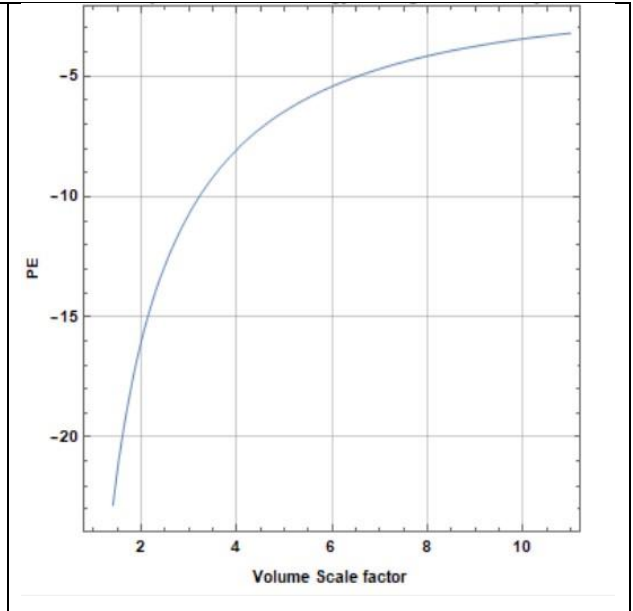 |
|---|---|
| Fig 14 : Kinetic energy vs Volume Scale Factor for negative viscosity | Fig 15: Potential energy vs Volume Scale Factor for negative viscosity. |

The kinetic energies of all the three cases remain positive and a decreasing function. In case of potential energy, we find an increasing function for without viscous and negative viscous cases. For the system with positive viscosity we find potential to be decreasing with time. So, we know that the potential energy should move from increasing nature to decreasing nature through a flat inflationary era. This mainly happens because the universe moves from virtual minima to real minima through a peak potential where the potential follows slow roll change. Thus, we may conclude that the universe should have either negative viscosity or no viscosity before inflation and positive viscosity after inflation. After inflation, the positive dissipation can bring the adequate-inadequate inflation and decelerated accelerating phases of universe which has to stop the high rate of expansion during inflation and start late time acceleration.

4. **Basics of finite time singularity problems and the Thermodynamics Energy conditions:**

For finite time singularity problems, we know the following types; [71, 49, 56, 68]

- Type I (known as Big Rip): At any finite time, $t = t_s$ we'll have $a \to \infty; \rho \to \infty; p \to \infty$
- Type II (known as Sudden Singularity): At any finite time, $t = t_s$ we'll have $a \to a_s; \rho \to \rho_s; p \to \infty$
- Type III: At any finite time, $t = t_s$ we'll have $a \to a_s; \rho \to \infty; p \to \infty$ ; it happens only in the EOS type of $p = -\rho - A \rho^\alpha$.

- Type IV: At any finite time, $t = t_s$ we'll have $a \to a_s$; $\rho \to \rho_s$; $p \to p_s$ Moreover, the Hubble rate and its first derivative also remain finite, but the higher derivatives, or some of these diverge. This kind of singularity mainly comes into play when $p = -\rho - f(\rho)$.

The thermodynamic energy conditions are basically derived from the well-known Raychaudhuri's equation. For a congruence of time-like and null-like geodesics, the Raychaudhuri equations are given in the following forms;

$$\frac{d\theta}{d\tau} = -\frac{1}{3}\theta^2 - \sigma_{\mu\nu}\sigma^{\mu\nu} + \omega_{\mu\nu}\omega^{\mu\nu} - R_{\mu\nu}u^\mu u^\nu \tag{17a}$$

And

$$\frac{d\theta}{d\tau} = -\frac{1}{3}\theta^2 - \sigma_{\mu\nu}\sigma^{\mu\nu} + \omega_{\mu\nu}\omega^{\mu\nu} - R_{\mu\nu}n^\mu n^\nu \tag{17b}$$

Where $\theta$ is the expansion factor, $n^\mu n^\nu$ is the null vector, and $\sigma_{\mu\nu}\sigma^{\mu\nu}$ and $\omega_{\mu\nu}\omega^{\mu\nu}$ are, respectively, the shear and the rotation associated with the vector field $u^\mu u^\nu$. For attractive gravity we'll have the followings;

$$R_{\mu\nu}u^\mu u^\nu \geq 0 \text{ and } R_{\mu\nu}n^\mu n^\nu \geq 0 \tag{17c}$$

To set some nomenclature the energy conditions of general relativity to be considered here are;

(i) Null energy condition (NEC)
(ii) Weak energy condition (WEC)
(iii) Strong energy condition (SEC)
(iv) Dominant energy condition (DEC)

So, for our matter-fluid distribution we may write this condition as follows;

- NEC = $\rho + p \geq 0$
- WEC = $\rho \geq 0$ and $\rho + p \geq 0$
- SEC = $\rho + 3p \geq 0$ and $\rho + p \geq 0$
- DEC = $\rho \geq 0$ and $-\rho \leq p \leq \rho$

The analytical discussion of the above points has been given below.

## 5. Energy conditions Investigation

From our derivations we see that the time varying function for scale factor, energy density and effective pressure can be expressed as follows.

$$a = a_0(a1 + bt)^m \tag{18a}$$

$$\rho_f = (6 - 2w_m)\lambda\rho_{m0}a_0^{-3(1+w_m)}(a1 + bt)^{-3(1+w_m)m} \tag{18b}$$

and

$$p_f = -2(1 - 3w_m)\lambda\rho_{m0}a_0^{-3(1+w_m)}(a1 + bt)^{-3(1+w_m)m} \tag{18c}$$

from the above functions we can easily conclude that all the finite time future singularities have been resolved from our calculations. Therefore, we need to discuss about the thermodynamics energy conditions.

### 5.1. Graphical representation of energy conditions for without viscous case

From further calculation we obtain Table- I for cosmic thermodynamic energy conditions without considering viscosity

**Table I**

| | |
|---|---|
| DEC | $\rho - \|p\| = 4(1 + w_m)\lambda\rho_{m0}a_0^{-3(1+w_m)}(a1 + bt)^{-3(1+w_m)m} \geq 0$ <br> So, $\infty \geq w_m \geq -1$ and $\lambda > 0$ or $-\infty < w_m < -1$ and $\lambda < 0$ |
| NEC | $\rho + p = 4(1 + w_m)\lambda\rho_{m0}a_0^{-3(1+w_m)}(a1 + bt)^{-3(1+w_m)m} \geq 0$ <br> So, $\infty \geq w_m \geq -1$ and $\lambda > 0$ or $-\infty < w_m < -1$ and $\lambda < 0$ |
| WEC | $\rho + p = 4(1 + w_m)\lambda\rho_{m0}a_0^{-3(1+w_m)}(a1 + bt)^{-3(1+w_m)m} \geq 0$ <br> So, $\infty \geq w_m \geq -1$ and $\lambda > 0$ or $-\infty < w_m < -1$ and $\lambda < 0$ <br> And <br> $\rho = (6 - 2w_m)\lambda\rho_{m0}a_0^{-3(1+w_m)}(a1 + bt)^{-3(1+w_m)m} \geq 0$ <br> So, $-\infty < w_m \leq 3$ and $\lambda > 0$ or, $w_m > 3$ and $\lambda < 0$ |
| SEC | $\rho + 3p = 16w_m\lambda\rho_{m0}a_0^{-3(1+w_m)}(a1 + bt)^{-3(1+w_m)m} \geq 0$ <br> So, $w_m \geq 0$ and $\lambda > 0$ or, $w_m < 0$ and $\lambda < 0$ <br> And <br> $\rho + p = 4(1 + w_m)\lambda\rho_{m0}a_0^{-3(1+w_m)}(a1 + bt)^{-3(1+w_m)m} \geq 0$ <br> So, $\infty \geq w_m \geq -1$ and $\lambda > 0$ or $-\infty < w_m < -1$ and $\lambda < 0$ |

In table I for Matter fields we have $w_m = 0$. So, the graphs for energy conditions will be as follows. Fig. 16 to 18.

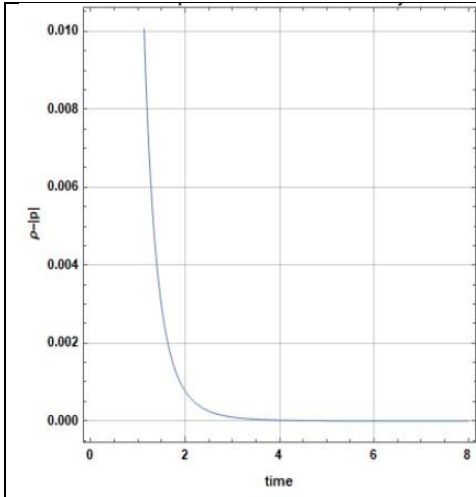 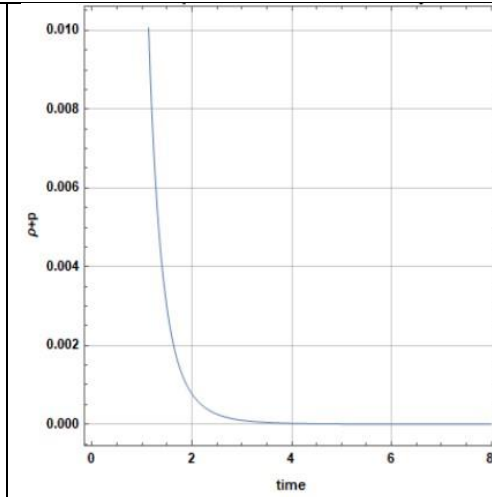 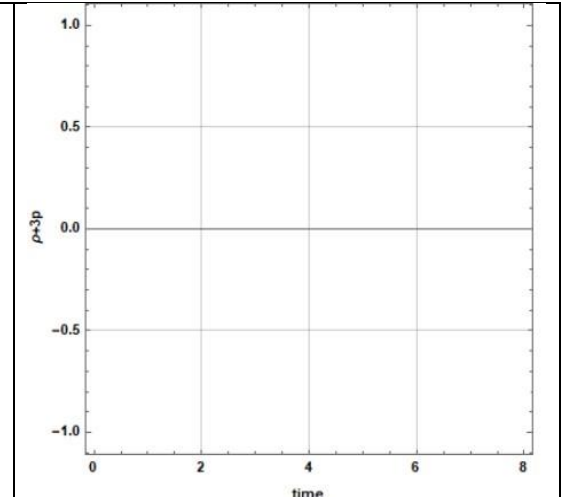

| Fig 16: DEC vs time without viscosity | Fig 17: WEC vs time without viscosity | Fig 18: SEC vs time without viscosity |

Here we see all the energy conditions are satisfied for without viscous case.

### 5.2. Graphical representation of energy conditions with viscosity

Now we'll apply the viscosity in the homogeneous linear fluid to introduce inhomogeneity. We'll use viscosity with its generalized format. [17]

We know that viscosity is an important factor to get highly negative pressure for hyperinflation and inflation (as $p' = p - 3\eta H$; $\eta$ = viscosity). Here $3\eta H = \eta_0(t)(3H)^{n+1} = 3^{\frac{n+1}{2}}\eta_0(t)(3H^2)^{\frac{n+1}{2}} = 3^{\frac{n+1}{2}}\eta_0(t)(\rho_m + \rho)^{\frac{n+1}{2}}$.

So, we get,

$p' = p - 3\eta H$

Or, $p' = p - 3^{\frac{n+1}{2}}\eta_0(t)(\rho_m + \rho)^{\frac{n+1}{2}}$ (19)

Now we have given the functional forms of the energy conditions with viscous cosmic fluid in Table- II
Here, $s = \frac{n+1}{2}$.

**Table II**

| DEC | $\rho - \|p'\| = 4(1 + w_m)\lambda\rho_{m0}a_0^{-3(1+w_m)}(a1 + bt)^{-3(1+w_m)m} + 3^s\eta_0(t)\big((6 - 2w_m)\lambda\rho_{m0} + \rho_{m0}\big)^s a_0^{-3(1+w_m)s}(a1 + bt)^{-3(1+w_m)sm} \geq 0$ <br> And |
|     | $\rho = (6 - 2w_m)\lambda\rho_{m0}a_0^{-3(1+w_m)} (a1 + bt)^{-3(1+w_m)m} \geq 0$ <br> So, $-\infty < w_m \leq 3$ and $\lambda > 0$ or, $w_m > 3$ and $\lambda < 0$ |
| NEC | $\rho + p' = 4(1 + w_m)\lambda\rho_{m0}a_0^{-3(1+w_m)} (a1 + bt)^{-3(1+w_m)m} - 3^s\eta_0(t)\big((6 - 2w_m)\lambda\rho_{m0} + \rho_{m0}\big)^s a_0^{-3(1+w_m)s} (a1 + bt)^{-3(1+w_m)sm} \geq 0$ |
| WEC | $\rho + p' = 4(1 + w_m)\lambda\rho_{m0}a_0^{-3(1+w_m)} (a1 + bt)^{-3(1+w_m)m} - 3^s\eta_0(t)\big((6 - 2w_m)\lambda\rho_{m0} + \rho_{m0}\big)^s a_0^{-3(1+w_m)s} (a1 + bt)^{-3(1+w_m)sm} \geq 0$ <br> And <br> $\rho = (6 - 2w_m)\lambda\rho_{m0}a_0^{-3(1+w_m)} (a1 + bt)^{-3(1+w_m)m} \geq 0$ <br> So, $-\infty < w_m \leq 3$ and $\lambda > 0$ or, $w_m > 3$ and $\lambda < 0$ |
| SEC | $\rho + 3p' = (16w_m)\lambda\rho_{m0}a_0^{-3(1+w_m)} (a1 + bt)^{-3(1+w_m)m} - 3^{s+1}\eta_0(t)\big((6 - 2w_m)\lambda\rho_{m0} + \rho_{m0}\big)^s a_0^{-3(1+w_m)s} (a1 + bt)^{-3(1+w_m)sm} \geq 0$ <br> And <br> $\rho + p' = 4(1 + w_m)\lambda\rho_{m0}a_0^{-3(1+w_m)} (a1 + bt)^{-3(1+w_m)m} - 3^s\eta_0(t)\big((6 - 2w_m)\lambda\rho_{m0} + \rho_{m0}\big)^s a_0^{-3(1+w_m)s} (a1 + bt)^{-3(1+w_m)sm} \geq 0$ |

In table II for Dark matter dominated universe $w_m = 0$. So, the graphs for energy conditions will be as follows. Fig 19 to 24.

**For positive viscosity:**

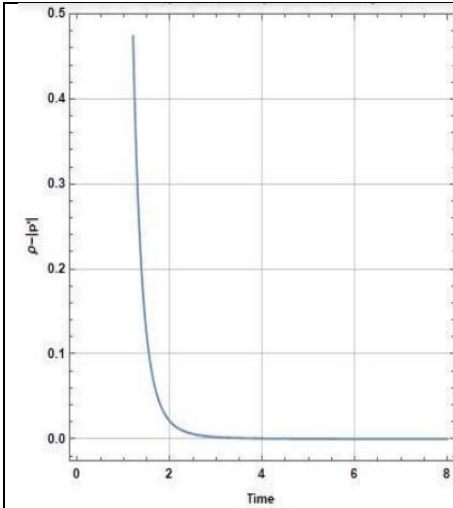

Fig 19: DEC vs time for positive viscosity

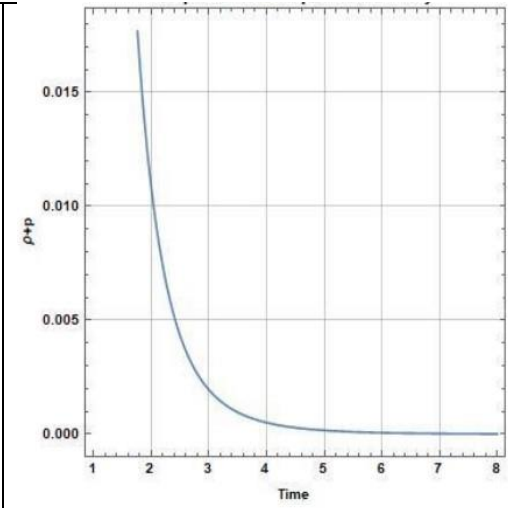

Fig 20: WEC vs time for positive viscosity

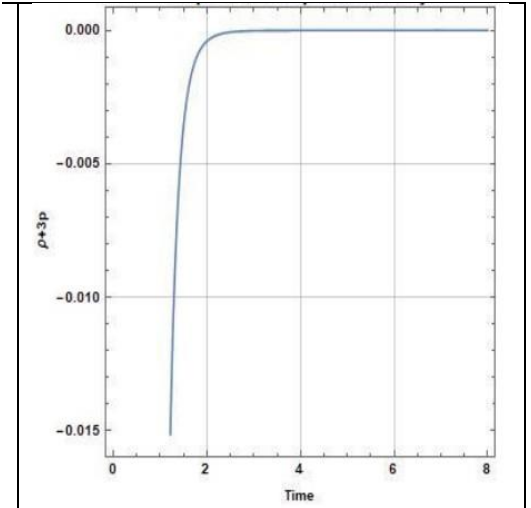

Fig 21: SEC time for positive viscosity

Here for positive viscosity SEC is not satisfied ($\eta > 0$).

**For negative viscosity:**

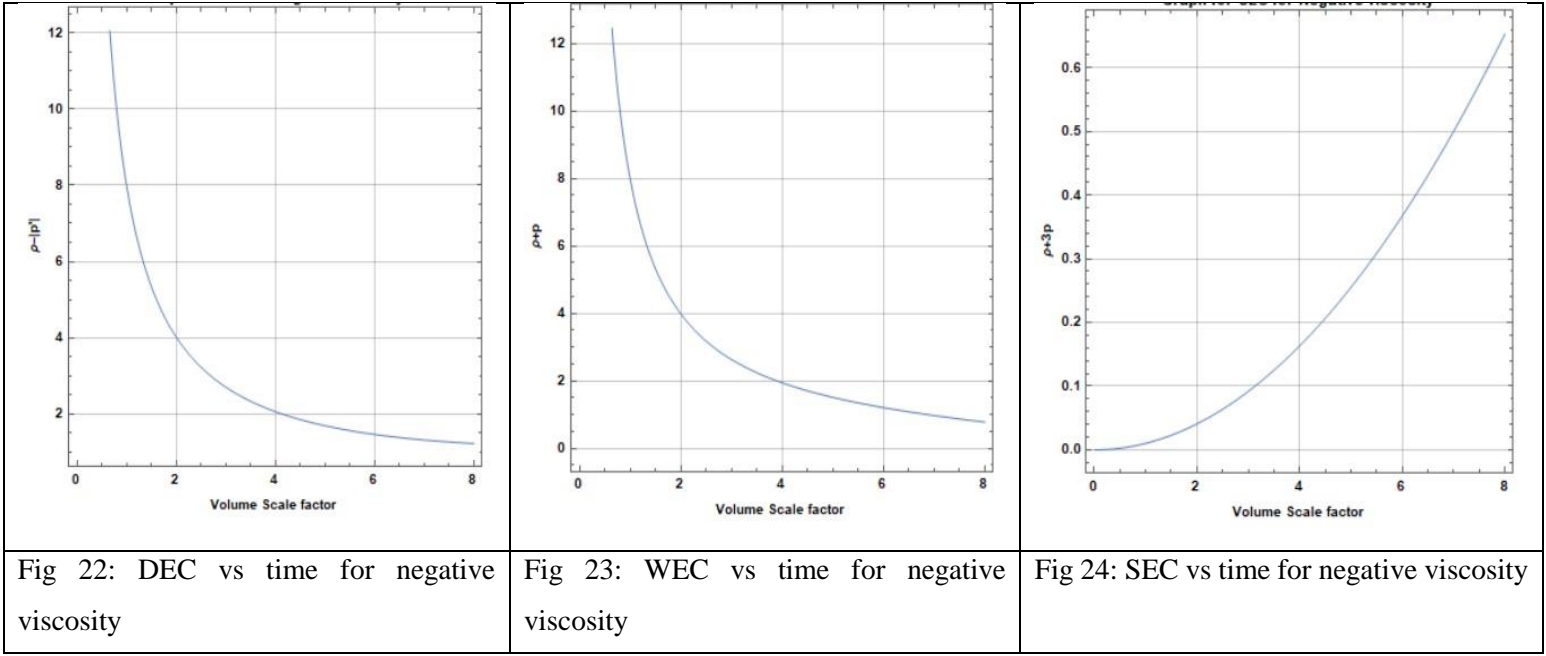

| Fig 22: DEC vs time for negative viscosity | Fig 23: WEC vs time for negative viscosity | Fig 24: SEC vs time for negative viscosity |

When we consider negative viscous coefficient ($\eta < 0$), all the energy conditions are satisfied.

## 6. Finite time future singularity and past singularity problems resolution

From the graphical representations of energy densities and pressures we can conclude that for any finite future time we cannot observe any singularities in energy densities and pressures. But we need to check the scale factor in those regions. The final form of the scale factor should be changed as the value and sign of the multiplicity and power parameters should be changed with the bouncing phase transition. The sign of time should also be changed. All those changes and conditions have been discussed below in the table. Although the original functional form of scale factor is $a(t) = a_0(a_1 + bt)^m$, the form will be changed after application of the sign of the parameters $m, b\ and\ a_1$.

From the assumed scale factor, we obtain Table- III.

**Table III**

| Cosmic phases | Functional Constants | Final form of the Function of scale factor |
|---|---|---|
| | | |

| Accelerated expansion | $m > 1; a_1 > 0$ and $b > 0$ for $t > 0$ | $a(t) = a_0(|a_1| + |bt|)^{|m|}$ |
| Decelerated expansion | $1 > m > 0; a_1 > 0$ and $b > 0$ for $t > 0$ | $a(t) = a_0(|a_1| + |bt|)^{|m|}$ |
| Contraction | $m < 0; a_1 > 0$ and $b < 0$ for $t < 0$ | $a(t) = \dfrac{a_0}{(|a_1| + |bt|)^{|m|}}$ |

From table III for accelerated expansion we have assumed $m > 1$ which gives $\ddot{a} > 0$. For decelerated expansion we have $1 > m > 0$ which provides $\ddot{a} < 0$.

At $t \to 0$ we assumed a cosmic bounce which is a transition point between contracting and expanding universe. This cosmic bounce provides a resolution of initial singularity problem. Before bounce the universe should pass through contracting phase for $t < 0$ and $b < 0$. So, we may conclude that the bouncing phase transition gives a change in the multiplicity parameter $b$ and power parameter $m$ from negative to positive values.

From the functions we can also observe that the universe should have the scale factor

$$a(t) = a_0(|a_1|)^m \qquad (20)$$

during bounce. Therefore, we can conclude that all types of singularities have been removed from scale factor, energy densities as well as pressures.

## 7. Thermodynamics for modified energy density and pressure with f (R, T) model and Inhomogeneity:

Here we will be introducing inhomogeneity through viscosity. The calculations for with and without viscosity will be done collaterally. [43-45] Here in this section we'll be discussing the internal energy, entropy, temperature and thermodynamic stability by following Chakraborty.et.al (2019) Panigrahi.et.al (2016) and Kar.et.al (2021). All the calculations have been done for both viscous and without viscous cosmology including only the modified gravity. Those results have been listed in Table -IV.

**Table IV**

| Thermodynamic variable | Without viscous case | With viscous case |
| --- | --- | --- |
| **Internal energy** | Here we have for $w_m = 0$, | Here also we have for $w_m = 0$, |
| | $w_{tot} = -\frac{2\lambda}{6\lambda+1}$, $p = -\left(\frac{\partial U}{\partial V}\right)_S$ and $\rho = \frac{U}{V}$<br><br>So, we get as follows.<br><br>$U = U_0 V^{\frac{2\lambda}{(6\lambda+1)}}$  Here $U_0 = V_0^{-\frac{2\lambda}{(6\lambda+1)}}$. | $w_{tot} = -\frac{2\lambda}{6\lambda+1} - 3^s \eta_0(t) \rho_{m0}^{s-1}(1+6\lambda)^{s-1} V^{-(s-1)}$,<br><br>$p = -\left(\frac{\partial U}{\partial V}\right)_S$ and $\rho = \frac{U}{V}$<br><br>So, we get as follows.<br><br>$U = U_0 V^{\frac{2\lambda}{(6\lambda+1)}} \exp\left(\frac{3^s \eta_0(t) \rho_{m0}^{s-1}(1+6\lambda)^{s-1}}{1-s} V^{-s+1}\right)$ Here<br><br>$U_0 = V_0^{-\frac{2\lambda}{(6\lambda+1)}} \exp\left(\frac{3^s \eta_0(t) \rho_{m0}^{s-1}(1+6\lambda)^{s-1}}{1-s} V_0^{-s+1}\right)$. |
| **Temperature**<br><br>(Here we know that the temperature and the dimensional definition for internal energy with temperature may be written as follows. [42, 43]<br><br>$T = \left(\frac{\partial U}{\partial S}\right)_V$ The differential equation for temperature can be written as follows.<br><br>$\frac{dT}{T} = -\frac{dV}{V} \frac{\partial p_{tot}}{\partial \rho_{tot}}$) | For viscous free case we may write;<br><br>$\frac{\partial p_{tot}}{\partial \rho_{tot}} = -\frac{2\lambda}{(6\lambda+1)}$<br><br>So, the temperature will be as follows.<br><br>$T = U_0 V^{\frac{2\lambda}{(6\lambda+1)}}$  Here $U_0 = V_0^{-\frac{2\lambda}{(6\lambda+1)}}$ | For viscous free case we may write;<br><br>$\frac{\partial p_{tot}}{\partial \rho_{tot}} = -\frac{2\lambda}{(6\lambda+1)} - 3^s s \eta_0(t) \rho_{m0}^{s-1}(1+6\lambda)^{s-1} V^{-(s-1)}$<br><br>So, the temperature will be as follows.<br><br>$T = U_0 V^{\frac{2\lambda}{(6\lambda+1)}} \exp\left(\frac{3^s s \eta_0(t) \rho_{m0}^{s-1}(1+6\lambda)^{s-1}}{1-s} V^{-s+1}\right)$ Here<br><br>$U_0 = V_0^{-\frac{2\lambda}{(6\lambda+1)}} \exp\left(\frac{3^s s \eta_0(t) \rho_{m0}^{s-1}(1+6\lambda)^{s-1}}{1-s} V_0^{-s+1}\right)$<br><br>The graphical representation of temperature will be similar to internal energy. |
| **Entropy**<br><br>(From the dimensional analysis of temperature and internal energy we may write as follows.<br><br>$[U] = [T][\Delta S]$) | For this case the dimensional analysis of internal energy will be as follows.<br><br>$[U] = [U_0][V]^{\frac{2\lambda}{(6\lambda+1)}}$<br><br>So, from equation (20) and (18) we can have, (As exponential function is dimensionless quantity)<br><br>$\Delta S = \frac{U_0}{T} V^{\frac{2\lambda}{(6\lambda+1)}}$  Or, $\Delta S = \frac{U_0}{T} V^{\frac{2\lambda}{(6\lambda+1)}}$ | For this case the dimensional analysis of internal energy will be as follows.<br><br>$[U] = [U_0][V]^{\frac{2\lambda}{(6\lambda+1)}}$<br><br>So, from equation (20) and (18) we can have,<br><br>$\Delta S = \frac{U_0}{T} V^{\frac{2\lambda}{(6\lambda+1)}}$  Or, $\Delta S = \frac{U_0}{T} V^{\frac{2\lambda}{(6\lambda+1)}}$ |
| **Thermodynamic stability**<br><br>(Here we have main three rules to have model stability thermodynamically with expanding | From the above derivations of pressure for viscous free fluid we may get,<br><br>$\left(\frac{\partial p_{tot}}{\partial V}\right)_S = \left(\frac{\partial p_{tot}}{\partial V}\right)_T = \frac{2\lambda \rho_{m0}}{V^2}$<br><br>So, we can observe for $\lambda > 0$ we have both $\left(\frac{\partial p_{tot}}{\partial V}\right)_S > 0$ and $\left(\frac{\partial p_{tot}}{\partial V}\right)_T > 0$.<br><br>Again, we can write, | From the above derivations of pressure for viscous free fluid we may get,<br><br>$\left(\frac{\partial p_{tot}}{\partial V}\right)_S = \frac{2\lambda \rho_{m0}}{V^2} - 3^s s \eta_0(t) \rho_{m0}^s (1+6\lambda)^s V^{-(s+1)}$<br><br>So, we can observe for $\lambda > 0$ we have $\left(\frac{\partial p_d}{\partial V}\right)_S < 0$ only when $3^s s \eta_0(t) \rho_{m0}^s (1+6\lambda)^s \gg 2\lambda \rho_{m0}$.<br><br>Now again, |

| universe. So, they are as follows. [44, 45] Rule I: $\left(\frac{\partial p_{tot}}{\partial V}\right)_S < 0$ Rule II: $\left(\frac{\partial p_{tot}}{\partial V}\right)_T < 0$ Rule-III: $c_v = T\left(\frac{\partial S}{\partial T}\right)_V > 0$ ) | $c_v = T\left(\frac{\partial S}{\partial T}\right)_V = -\frac{U_0}{T} V^{\frac{2\lambda}{(6\lambda+1)}}$ So, for without viscosity case we observe none of the thermodynamics stability rules are being satisfied. We make an attempt to satisfy all the conditions by considering viscous loss. | $p_{tot} = -2\lambda \rho_{m0} V^{-1} - \frac{\rho_{m0}(1+6\lambda)(1+s)}{s}\Big[\ln T - \ln U_0 - \frac{2\lambda}{(6\lambda+1)} \ln V\Big] V$ So, we get, $\left(\frac{\partial p_{tot}}{\partial V}\right)_T = \frac{2\lambda \rho_{m0}}{V^2} - \frac{\rho_{m0}(1+6\lambda)(1+s)}{s}\Big[\ln T - \ln U_0 - \frac{2\lambda}{(6\lambda+1)} \ln V\Big] + \frac{2\lambda \rho_{m0}(1+6\lambda)(1+s)}{s(6\lambda+1)}$ Or, $\left(\frac{\partial p_{tot}}{\partial V}\right)_T = \frac{2\lambda \rho_{m0}}{V^2} - 3^s s \eta_0(t) \rho_{m0}^s (1 + 6\lambda)^s V^{-(s+1)} + \frac{2\lambda \rho_{m0}(1+6\lambda)(1+s)}{s(6\lambda+1)}$ Here also to satisfy the stability rule II we must have $3^s s \eta_0(t) \rho_{m0}^s (1 + 6\lambda)^s \gg \left(2\lambda \rho_{m0} + \frac{2\lambda \rho_{m0}(1+6\lambda)(1+s)}{s(6\lambda+1)}\right)$. Again, we can write, $c_v = T\left(\frac{\partial S}{\partial T}\right)_V = -\frac{U_0}{T} V^{\frac{2\lambda}{(6\lambda+1)}}$ We observe that after considering viscous loss the first two rules of thermodynamic stability are satisfied but the specific heat change remains negative. Therefore, for model stability we must have $U_0 < 0$. We can also consider multiple fluid system where we'll be allowed to consider $U_0 > 0$ or $c_v < 0$. |

So, using table IV the graphs of internal energy variation can be plotted as follows. Fig. 25 to 27.

| 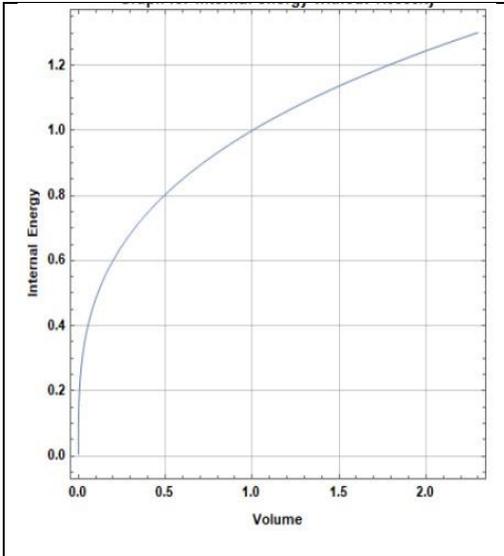 | 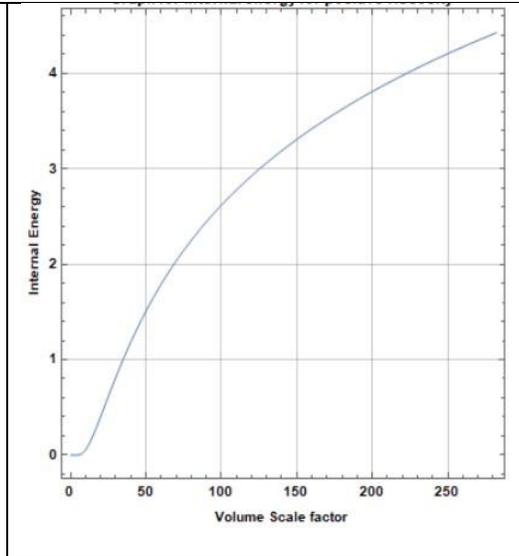 | 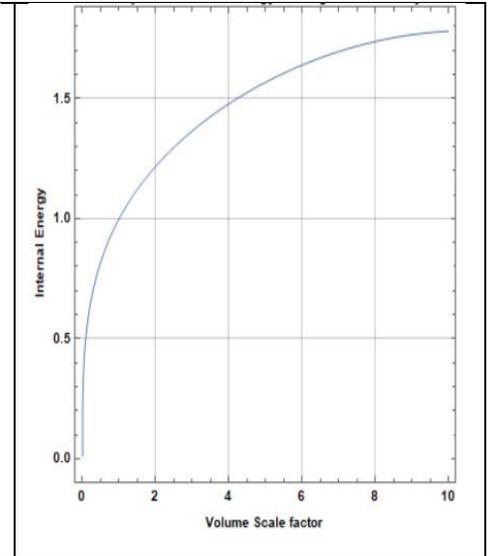 |
|---|---|---|
| Fig 25: Internal energy vs Volume Scale Factor without viscosity | Fig 26: Internal energy vs Volume Scale Factor for positive viscosity | Fig 27: Internal energy vs Volume Scale Factor for negative viscosity |

From the calculations in table IV we found that the entropy change, constant volume specific heat/ heat capacity and temperature all of them varies in a similar fashion. We are plotting single graph for each variable for all those conditions in fig. 28 to 31.

| 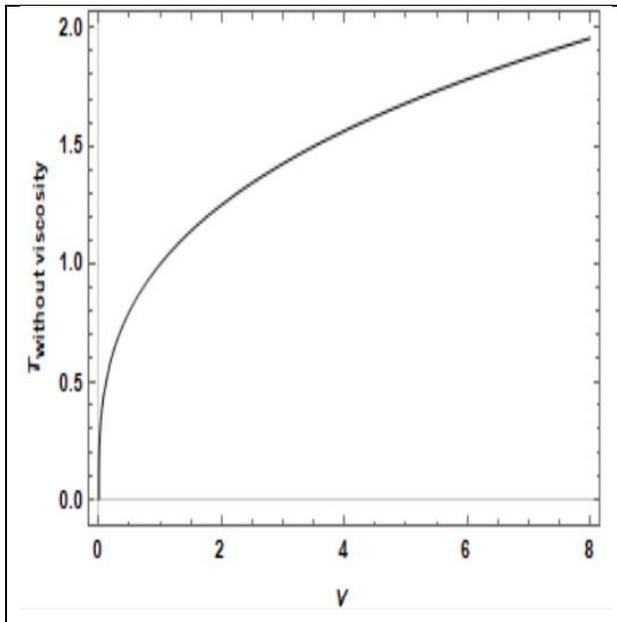 | 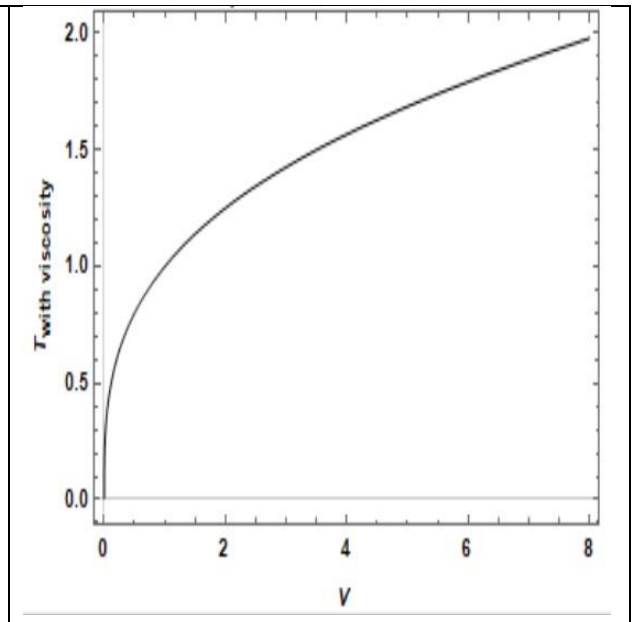 |
|---|---|
| Figure 28: Temperature for without viscosity | Figure 29: Graph for Temperature with viscosity |

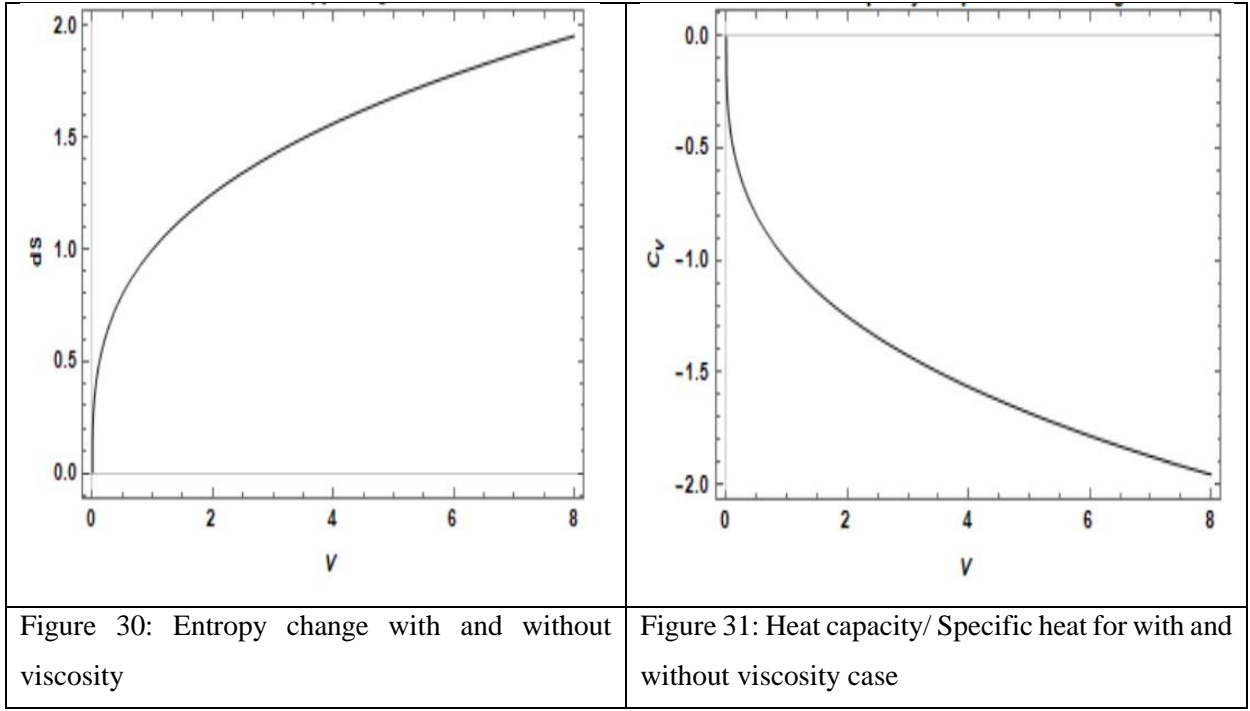

| Figure 30: Entropy change with and without viscosity | Figure 31: Heat capacity/ Specific heat for with and without viscosity case |

In the above calculations we found two main problems and they are negative heat capacity problem and negative viscosity problem. We know that none of them exist with single type fluid cosmic model since they don't obey Generalized second law of thermodynamics and we need to introduce multiple system to resolve these problems through interaction between the fluid components. We'll discuss this problem in the section below.

## 8. Introduction of multiple fluid system and solution of negative heat capacity problem:

In the calculations of table IV, we found positive change of entropy and negative heat capacity. [65] This negative heat capacity is one kind of a paradox in physics especially found in astronomy and astrophysics. Here we found a similar paradox while discussing cosmic evolution. This paradox has been resolved several times in literature for astronomical and astrophysical problems. Here we have taken an interesting step to resolve such paradox and i.e. we have introduced the multiple fluid system in our calculation. We have used two special type of energy transition cycles that provides an exact definition of cosmic inflation, decelerated expansion, graceful exit, and reheating as well as present state of accelerated expansion [see appendix]. In the discussion of thermodynamics energy conditions and thermodynamics stability analysis for expanding universe we found some difficulties when we used viscosity. Even we found that the viscous free fluid is inefficient to express the expanding model stability thermodynamically. So, such ambiguity has also been resolved with this energy transition cycle in multiple fluid system. [65, 66]

### 8.1. Theoretical presentation of resolution of this problem:

We consider two collateral system named $S_1$ and $S_2$. We assume that these two systems will remain connected through energy transition. Those cycles have been listed below with their block diagrams.

| | |
|---|---|
| Block diagram 1: Cycle I | 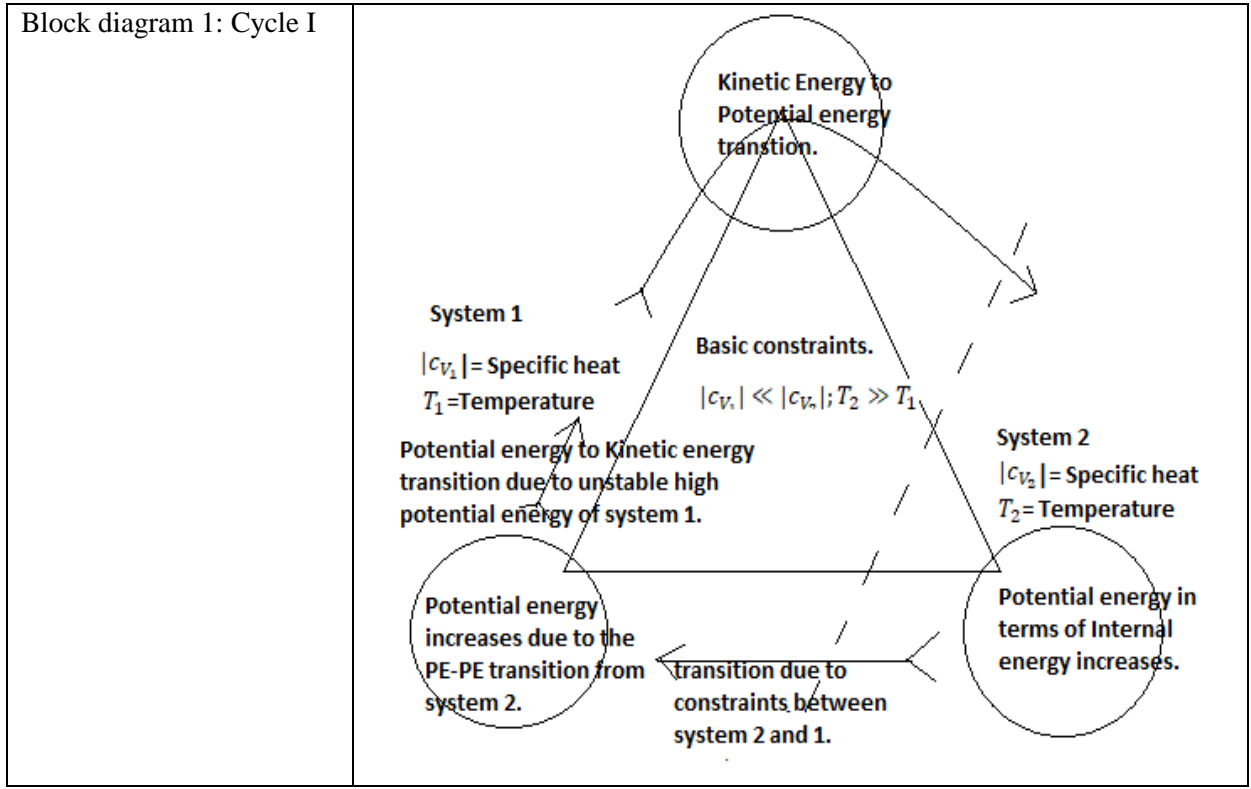 |

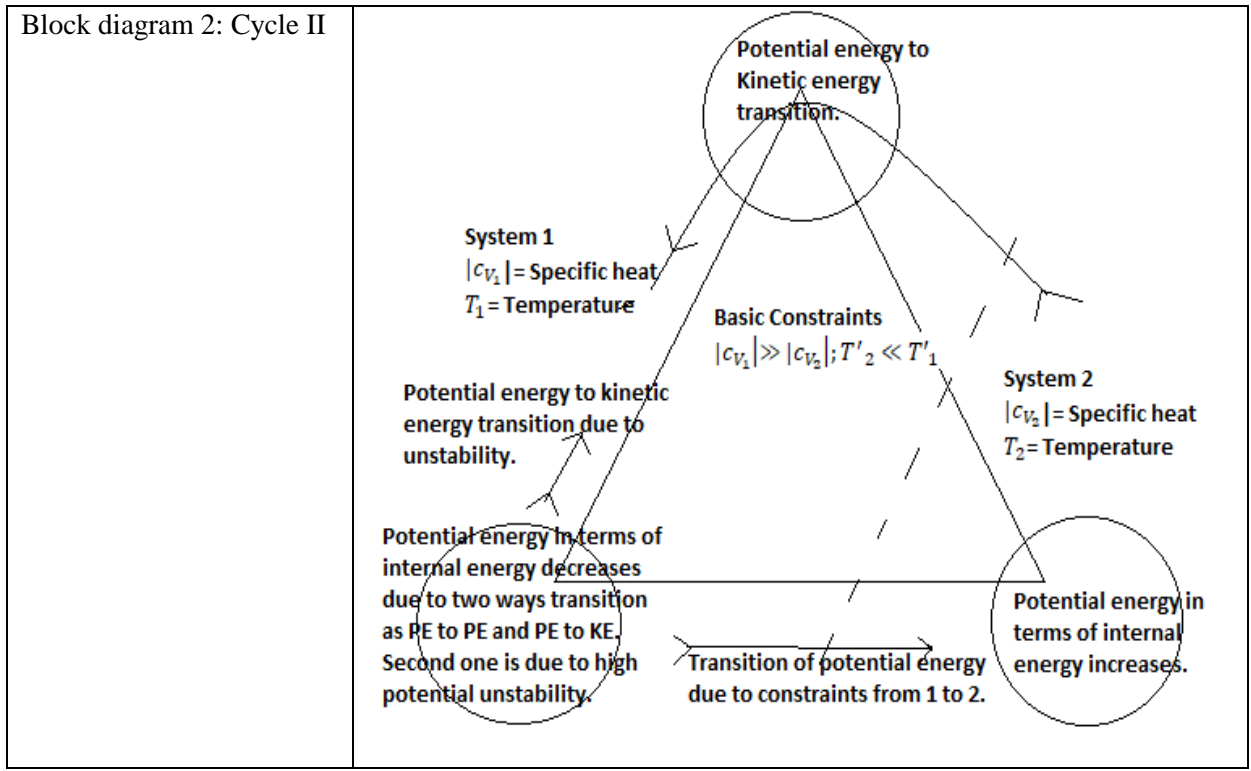

Block diagram 2: Cycle II

In this discussion we have followed some basic assumptions in terms of thermodynamics. From dimension analysis we have considered the internal energy as $[U] = [T][S]$. This is different from the usual form of first law of generalized thermodynamics. In functional form this relation can be considered as $TdS = dU - SdT$. Here $TdS$ = Change of total energy = $dQ$, $dU$ = Change of potential energy in terms of Internal energy and $-SdT$ = Change of energy due to change of temperature that measures the change of kinetic energy = $d(KE)$. So, overall the relation will become $dQ = dU + d(KE) = d(PE) + d(KE)$.

**Cycle I:** Here this cycle follows the basic constraints relations as → $|c_{v_1}| \ll |c_{v_2}|$ and $T_2 \gg T_1$. Another constraint for this cycle is the negative heat capacity i.e. the temperature will decrease with increase in internal energy or potential energy $\left(\frac{\partial U}{\partial T} < 0\right)$. (Block diagram 1)

Due to change of potential energy or internal energy with cosmic evolution, the temperature increases and thus the kinetic energy also increases. This can be represented as the potential energy to kinetic energy transfer in system 1. Now the constraints tell us that for same change in internal potential energy, the temperature change in system 2 should be less than that of system 1 and thus the temperature difference doesn't vanish. This should cause lower kinetic energy in system 2, and thus we can find an energy transition from kinetic energy in system 1 to potential energy in system 2 (In every case of kinetic energy to kinetic energy transition the gainer should increase its potential energy first, and then this increased potential energy should then convert into kinetic energy). The increased potential energy in terms of internal energy will then transfer into potential energy of system 1 due to temperature difference (this is one kind of internal interaction between system 1 and system 2). Thus, a cycle of energy transition will be completed with the condition that the energy transition for KE to PE transition is less than that for PE to PE transition. So, there will have a resultant heat loss in system 1 and resultant increase in temperature in system 2.

For system 1 this fact will be similar to one directional KE loss and we have mentioned it as equivalent to positive viscosity effect with s = 2 or s > 2. This cycle will run from $w_{EOS} < -1$ to $w_{EOS} = -\frac{1}{3}$. For $w_{EOS} = -1$ we get adequate inflation or hyperinflation and for $-\frac{1}{3} > w_{EOS} > -1$ universe should face inadequate inflation. [57]

The cycle I will stop when the temperature of system 2 will reach its critical point and when temperatures and resultant lost heat will cause a spontaneous symmetry breaking in universe supersymmetric fields. This spontaneous symmetry break should bring a second order phase transition with discontinuous change in heat capacity and nature of viscosity. Thus, Cycle II will start.

The resultant heat loss through the entire cycle will produce radiation pressure component in the Friedmann equation that pushes the SEC into positive values (As radiation pressure is a positive pressure). Thus, All the thermodynamics energy conditions for attractive gravity are resolved with accelerated expansion nature of universe. [72]

The cycle runs with positive change of entropy. Thus, both the first law and second law of thermodynamics have been satisfied in this cycle along with the resolution of negative heat capacity.

**Cycle II:** Here this cycle follows the basic constraint relations as → $|c_{v_1}| \gg |c_{v_2}|$ and $T'_2 \ll T'_1$. Another constraint for this cycle is the negative heat capacity i.e. the temperature will decrease with increase in internal energy or potential energy $\left(\frac{\partial U}{\partial T} < 0\right)$ for system 1 and positive heat capacity for system 2. After the second order phase transition in cycle I the heat capacity for system 1 got discontinuous change but keeps negative value and for system 2 got discontinuous change with positive value. (Block diagram 2)

Due to change of potential energy or internal energy with cosmic evolution, the temperature increases and thus the kinetic energy will also increase (similar phenomena as cycle I). As in this cycle the constraints are different and that's why the increase in internal energy increases the temperature of system 2 but decreases the temperature of system 1. The increased potential energy in terms of internal energy will convert to its kinetic energy and due to higher change in temperature ($\Delta T'$), this kinetic energy will be directly transferred into system 1 as kinetic energy (due to sudden instability). So, overall, we can say that there is a potential energy to kinetic energy transition from system 2 to system 1 (As the KE produced from PE in system 2 will convert into KE of system 1 and thus the PE of system 2 converts into KE of system 1). Again, due to the constraints relation a potential energy to potential energy transition in terms of internal energy will happen from system 1 to system 2. This decrease in potential energy may cause increase in temperature in system 1 and thus the energy transition continues. At the end of this cycle the system 1 will face a resultant heat loss. [58]

For cycle II the system 1 will have an increase in energy in terms of kinetic energy (due to the PE to KE transition from system 2 to system 1). This is equivalent to negative viscosity effect for system 1. This cycle will run from $w_{EOS} = -\frac{1}{3}$ to $w_{EOS} = \frac{1}{3}$. The universe should face decelerated expansion in this cycle. [102]

The resultant heat loss and the continuous increasing temperature of system 1 will bring it to a critical point where the universe will face another spontaneous symmetry breaking. This will cause a second order phase transition and a discontinuous change in heat capacities in those two systems. Thus, cycle II will stop with graceful exit. The cycle I will start with the reheating phase of radiation dominated universe.

The resultant heat loss through the entire cycle will produce radiation pressure component in the Friedmann equation that pushes the SEC into positive values (As radiation pressure is a positive pressure). Thus, All the thermodynamics energy conditions for attractive gravity are resolved with accelerated expansion nature of universe. [60]

The cycle runs with positive change of entropy. Thus, both the first law and second law of thermodynamics has been obeyed in this cycle along with the resolution of negative heat capacity problems in our research.

**Starting of first Cycle I at the beginning of universe:**

At the very beginning of the universe expansion there must have some phase transition and a transition from cycle II to Cycle I. But it is impossible to have another expanding phase before the start of universe expansion. Thus, we may conclude, that there must have been some contracting phase of universe with negative change in entropy ($\Delta S < 0\ as\ volume\ decreases$). In this phase the cycle II should undergo third order phase transition with continuous change in $\Delta S$, volume as well as in heat capacity. After this discontinuous change in entropy the universe got into expanding phase (as $\Delta S\ becoming > 0$). The discontinuous change in volume provides the minimum volume of universe that keeps the shape of universe from having point singularity with quantum degeneracy pressure. This represents an idea of bouncing nature of universe. It also helps to solve early universe point singularity problems. [61-63]

Hence, we can now establish the viability of those cycles theoretically as well as thermodynamically.

**Cosmic bounce phase transitions**

In table III we have observed that to solve past singularity and initial singularity the value of power parameter $m$ and multiplicity parameter $b$ must change from negative to positive values. This transition can be explained by a cosmic bounce phase transition that changes the contracting universe to an accelerated universe. According to our assumed scale factor this transition happened at time $t \to 0$. Since we observe from graphs 28 to 31 that first order derivative of entropy and specific heat both are continuous at $t \to 0$, we can conclude that this transition is a third order phase transition. This phase transition happens between the past time cycle II ($t < 0$) to pre-inflationary cycle I ($t > 0$).

**Inflationary phase transition**

We get inflation from the transition between cycle I to cycle II during which $w_{EOS} \to -1$. For inflation start we need very high potential that is produced by negative viscosity (as observed in figure 15). The early universe high negative pressure can cause this inflation. The late time decrease in the negativity of pressure may cause to bring the adequate and inadequate inflationary phases after inflation and cause the decelerated expansion phase.

| 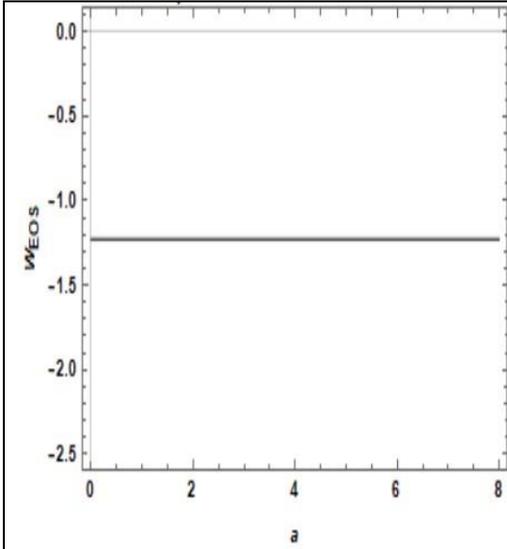 | 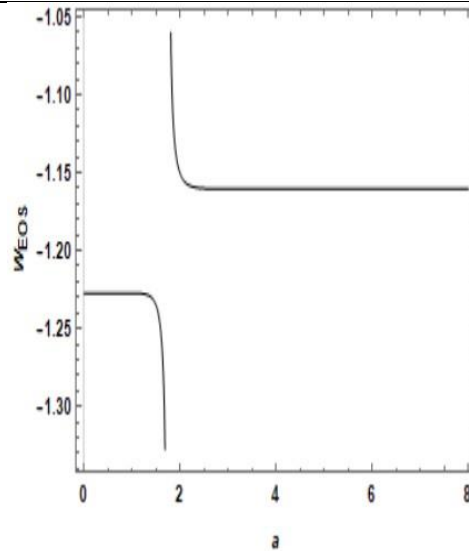 | 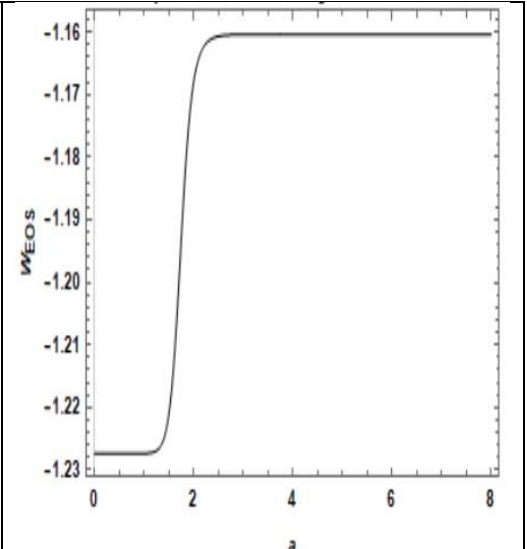 |
|---|---|---|
| Figure 32: $w_{EOS}$ for without viscosity | Figure 33: $w_{EOS}$ for with positive viscosity | Figure 34: $w_{EOS}$ for with negative viscosity |

Here we know that $w_{EOS} = \frac{KE-PE}{KE+PE}$. From this expression we can observe that before inflation the fluid behaves as phantom like system and after inflation it becomes quintessence like fluid.

From the nature of $w_{EOS}$ that can be derived from graphs of Kinetic energy and potential energy (fig.10 to 15 and 32 to 34) we can conclude that for $w_{EOS} < -1$ negative viscosity is dominant but for $w_{EOS} \geq -1$ (Inflation phase and late time acceleration) positive viscosity is dominant. Therefore, there is a transition from negative viscous fluid to positive viscous fluid during the transition from cycle I to Cycle II.

**Decelerated expansion, graceful exit and reheating**

As previously discussed cycle II denotes decelerated expansion of universe and this is caused by positive viscosity. Further transition from cycle II to Cycle I will cause graceful exit and reheating. Reheating brings the start of radiation dominated universe. These processes are followed by the late time acceleration that we witness now.

**Resolution of negative viscosity problem**

In general, we know that viscosity causes dissipation of energy. But for negative viscous fluid it causes gain of energy. This is thermodynamically not viable for a single fluid since there is no other source to gain energy from, for an isolated system. In our discussion, we have introduced the multiple fluid concept where a single component can have negative viscosity that causes it to gain energy which is supplied by the other component. Hence, overall the effective system obeys the laws of thermodynamics.

**Present universe temperature problem:**

In cycle I we see there should be a continuous increase in temperature $T_2$ which opposes the present universe experimental results and i.e. the universe should have a decreasing temperature change. So, we get another contradiction in our theory. This problem can also be resolved as follows.

This problem can be solved with introducing a third system with scalar field. The continuously increased temperature will cause the increase in scalar field potential for that third system. This phenomenon may cause present universe to have decreasing temperature and also be DE dominated. So, for cycle I there will not be any direct heat loss. The resultant heat lost in cycle II before starting reheating will cause the necessary heat in radiation dominated universe. This phenomenon may happen when we include the scalar field to fluid interaction. [64] [see section 9.2]

### 8.2. Mathematical presentation of those two cycles:

In accordance to our multiple fluid system we try to solve the above problem mathematically. According to our convention let do the calculations as follows.

**Cycle I:**

For those two systems the final energy conservation equations can be written as follows.

$$\dot{\rho}_1 + 3H(\rho_1 + p"_1) = 0 \tag{21a}$$

And

$$\dot{\rho}_2 + 3H(\rho_2 + p"_2) = 0 \tag{21b}$$

Where $p"_1 = p_1 - \eta_0(t)(3H)^{n+1} + \frac{c_{V_2}\Delta T_2}{3H}$ and $p"_2 = p_2 + \eta_0(t)(3H)^{n+1} - \frac{c_{V_2}\Delta T_2}{3H}$.

If we say $\eta_0(t)(3H)^{n+2} = Q_1$ and $c_{V_2}\Delta T_2 = Q_2 = c_{V_1}\Delta T_1$ then we may write as follows.

$$\dot{\rho}_1 + 3H(\rho_1 + p_1) = Q_1 - Q_2 \tag{21c}$$

And

$$\dot{\rho}_2 + 3H(\rho_2 + p_2) = -Q_1 + Q_2 \tag{21d}$$

So, we can observe the equations (22c) and (22d) will represent the cycle I energetically and mathematically. Here from the constraints and assumptions taken in cycle I we may say that $Q_2 - Q_1 = \Delta Q > 0$ and the pressure that helps to obey energy conditions for the universe may be written as $p"_1 = p_1 + \frac{\Delta Q}{3H}$. Here $\eta_0(t)$ is positive coefficient of viscosity. Here $|Q_2| \gg |Q_1|$. $Q_1$ is the KE to PE transition and $Q_2$ is the PE to PE transition according to theoretical discussion.

**Cycle II:**

For those two systems the final energy conservation equations can be written as follows.

$$\dot{\rho}_1 + 3H(\rho_1 + p"_1) = 0 \tag{22a}$$

And

$$\dot{\rho}_2 + 3H(\rho_2 + p"_2) = 0 \tag{22b}$$

Where $p"_1 = p_1 + \eta'_0(t)(3H)^{n+1} - \frac{c'_{V_2}\Delta T'_2}{3H}$ and $p"_2 = p_2 - \eta'_0(t)(3H)^{n+1} + \frac{c'_{V_2}\Delta T'_2}{3H}$.

If we say $\eta'_0(t)(3H)^{n+2} = Q'_1$ and $c'_{V_2}\Delta' T_2 = Q'_2 = c_{V_1}\Delta T'_1$ then we may write as follows.

$$\dot{\rho}_1 + 3H(\rho_1 + p_1) = -Q'_1 + Q'_2 \tag{22c}$$

And

$$\dot{\rho}_2 + 3H(\rho_2 + p_2) = Q'_1 - Q'_2 \tag{22d}$$

So, we can observe the equations (23c) and (23d) will represent the cycle I energetically and mathematically. Here from the constraints and assumptions taken in cycle II we may say that $Q'_1 - Q'_2 = \Delta Q' > 0$ and the pressure that helps to obey energy conditions for the universe may be written as $p''_1 = p_1 + \frac{\Delta Q'}{3H}$. Here $\eta'_0(t)$ is positive coefficient of viscosity. Here $|Q'_2| \ll |Q'_1|$. $Q'_1$ is the PE to KE transition and $Q'_2$ is the PE to PE transition according to the discussion.

In both the cycles the energy conservation equations for second fluid system contained modified pressure but this is written just to represent the energy analysis. The second fluid system may be viscous free or inhomogeneous.

### 8.3. Identification of those binary fluid system (S1 and S2) through specified EOS and conservation equations:

According to our earlier discussion the binary fluid system may be introduced into the Friedmann equations as follows.

$$3H^2 = \rho_m + \rho_{f_1} + \rho_{f_2} + \rho_{\varphi_1} \tag{23a}$$

And

$$2\dot{H} + 3H^2 = -p_m - p_{f_1} - p_{f_2} - p_{\varphi_1} \tag{23b}$$

Where $\rho_{f_1}$ and $\rho_{f_2}$ are the energy densities for those two fluids $S_1$ and $S_2$ respectively. The second fluid belongs to the EOS as follows in equation (25). The scalar field will help to discuss the present day DE dominated universe. Here we are discussing without interaction case.

$$p_{f_2} = A\rho_{f_2} + Ba^\alpha \tag{24}$$

So, the minimally coupled action will be as follows.

$$S = \int d^4x\, a^3(t) [\sqrt{-g} f(R,T) + L_m] + S_{\varphi_1} \tag{25}$$

Where $S_{\varphi_1} = \int d^4x\, a^3(t) \left[ T(\varphi_1) \sqrt{1 - \frac{\dot{\varphi}_1^2}{T(\varphi_1)}} + V(\varphi_1) - T(\varphi_1) \right]$.

The equation of conservations can be written as follows.

$$\dot{\rho}_m + 3H(\rho_m + p_m) = 0 \tag{26a}$$

$$\dot{\rho}_{f_1} + 3H(\rho_{f_1} + p''_{f_1}) = 0 \tag{26b}$$

$$\dot{\rho}_{f_2} + 3H(\rho_{f_2} + p''_{f_2}) = 0 \tag{26c}$$

$$\dot{\rho}_{\phi_1} + 3H(\rho_{\phi_1} + p_{\phi_1}) = 0 \tag{26d}$$

For the second system of inhomogeneous fluid the internal energy, energy density, pressure, entropy and heat capacity can be written as follows (considering that fluid is an isolated system).

**Internal energy:** $U = -\frac{3B}{\alpha+3A+3} V^{\frac{\alpha}{3}+1} + constant * V^{-A}$ (27a)

**Energy density:** $\rho_{f_2} = -\frac{3B}{\alpha+3A+3} V^{\frac{\alpha}{3}} + constant * V^{-(A+1)}$ (27b)

**Pressure:** $p_{f_2} = B\left[1 - \frac{3}{\alpha+3A+3}\right] V^{\frac{\alpha}{3}} + A * constant * V^{-(A+1)}$ (27c)

**Entropy:** $S = \frac{constant}{T} V^{-A}$ (27d)

**Heat capacity:** $c_{v_2} = T\left(\frac{\partial S}{\partial T}\right)_V = -\frac{constant}{V^A} \frac{1}{T}$ (27e)

Although the fluids in that binary fluid system behave as energetically open systems in canonical as well in grand canonical ensemble. For energetically isolated system like microcanonical ensemble the negative heat capacity paradox can't be resolved as in this case negative entropy is not a paradox instead it is a reality (according to virial theorem) [71, 72],

9. **Generalized solution of scale factor and inhomogeneous fluid density from minimally coupled DBI Essence scalar field:**

Previously we have proceeded to calculate all the variables by assuming the scale factor of the form $a = a0(a1 + bt)^m$. But this was viable only for single component fluid system of universe. From our previous calculation, we observed that we need multiple fluid system to explain the evolution of universe. Therefore, we tried to find a common definition of scale factor which may satisfy the new system including Cycle I and Cycle II. Here we'll find the generalized definition of scale factor that is reconstructed with multiple fluid system.

From the equations (26a), (26b), (26c) and (26d) we can rewrite them as follows.

$$\dot{\rho}_m + 3H(\rho_m + p_m) = 0 \tag{28a}$$

$$\dot{\rho}_f + 3H(\rho_f + p_f) = 0 \tag{28b}$$

$$\ddot{\phi}_1 + 3H\dot{\phi}_1 + \frac{\partial V}{\partial \phi_1} = 0 \tag{28c}$$

The binary fluid terms have been added to make the conservation equation modified pressure free. In equation (28c) we used quintessence scalar field theory instead of DBI essence theory to simplify our calculation.

The solution of equation (28a) and (28b) can be written as follows. (using $w_m = 0$)

$$\rho_m = \rho_{m0} a^{-3} \tag{29a}$$

And

$$\rho_f = \rho_{f0} a^{-3(1+w_f)} \tag{29b}$$

From equation (28c) we get the solution of scalar field density can be written as follows. (considering potential as a function of scalar field kinetic energy)

$$\rho_{\phi_1} = \rho_{\phi_1 0} a^n \tag{29c}$$

So, from Friedmann first equation we get the scale factor relation as follows.

$$\frac{\dot{a}}{a} = \sqrt{\rho_{m0} a^{-3} + \rho_{f0} a^{-3(1+w_f)} + \rho_{\phi_1 0} a^n}$$

Or, $\int \dfrac{da}{\sqrt{\rho_{m0} a^{-1} + \rho_{f0} a^{-3\left(\frac{1}{3}+w_f\right)} + \rho_{\phi_1 0} a^{n+2}}} = kt + k_1$

$$\tag{29d}$$

Thus, from this integration we can get the time variation of scale factor for our non-interacting system.

If we use DBI essence model in equation (28c) we may get,

$$\ddot{\phi}_1 - \frac{3T'(\phi_1)}{2T(\phi_1)} \dot{\phi}_1^2 + T'(\phi_1) + \frac{3}{\gamma^2}\frac{\dot{a}}{a}\dot{\phi}_1 + \frac{1}{\gamma^3}[V'(\phi_1) - T'(\phi_1)] = 0 \tag{30a}$$

So, the solution for energy density will be,

$$\rho_{\phi_1} = 3 \log \frac{a}{c} \tag{30b}$$

So, the scale factor solution will be as follows.

$$\int \frac{da}{\sqrt{\rho_{m0} a^{-1} + \rho_{f0} a^{-3\left(\frac{1}{3}+w_f\right)} + 3a^2 \log \frac{a}{c}}} = k_2 t + k_3 \tag{30c}$$

This provides similar reconstruction of scale factor with DBI essence scalar field.

## 10. Basics of Stability analysis with cosmological perturbation and state finder diagnostics:

Here in this section we'll discuss the basics of model stability with cosmological perturbation as well as the characterization of our research w.r.t state finder diagnostics. The main equation to describe the cosmological perturbation can be written as follows. [49-56]

$$\ddot{\phi}_k + \alpha \dot{\phi}_k + \mu^2 \phi_k + c_s^2 \frac{k^2}{a^2} \phi_k = 0 \tag{31}$$

Where,

$$\alpha = 7H + \frac{2\partial V(\phi)}{\partial \phi} \tag{31a}$$

$$\mu^2 = 6H^2 + 2\dot{H} + \frac{\frac{2H\partial V(\phi)}{\partial \phi}}{\dot{\phi}} \tag{31b}$$

$$c_s^2 = \frac{\dot{p}_f}{\dot{\rho}_f} \tag{31c}$$

For being stable with such cosmological perturbation the parameters should be acquired positive values for the respective model.

There are several models which are efficient to explain the accelerated universe properly. To differentiate those models one from another, a special diagnostics method has been proposed by Sahni (2003). This diagnostic is basically an analysis of one pair of parameters {r, s} where the trajectories for each model on r-s plane help in differentiation of these models. The mathematical format of those parameters may be written as follows.

$$r = \frac{\dddot{a}}{aH^3} \tag{32a}$$

$$s = \frac{r-1}{3\left(q-\frac{1}{2}\right)} \tag{32b}$$

Where q is the deceleration parameter which has the form as follows.

$$q = -\frac{a\ddot{a}}{\dot{a}^2} \tag{32c}$$

The above mathematical forms of the state finder parameters and deceleration parameter can be rewritten w.r.t pressure and density as follows.

$$r = 1 + \frac{9}{2(\Sigma \rho_i)} \left[ \Sigma \frac{\partial p_i}{\partial \rho_i}(p_i + \rho_i) \right] \tag{32d}$$

$$s = \frac{1}{\Sigma p_i} \left[ \Sigma \frac{\partial p_i}{\partial \rho_i}(p_i + \rho_i) \right] \tag{32e}$$

$$q = \frac{1}{2} + \frac{3}{2}\left[\frac{\sum p_i}{\sum \rho_i}\right] \tag{32f}$$

For accelerating universe, the deceleration parameter should have negative value and for deceleration it should have positive value.

### 11. Analysis and graphical representations for those parameters:

In the above calculations most of the important results and solutions have been done with the modified pressure for viscous fluid. Here we'll represent our analysis for all those cases including without viscous fluid as well as fluid with both negative and positive viscosity.

#### 11.1. State finder parameters diagnostics:

From the equations (32d), (32e) and (32f) we can write as follows.

$$r = 1 + \frac{9}{2(\rho_{f_1}+\rho_{f_2})}\left[\frac{\partial p_{f_1}}{\partial \rho_{f_1}}(p_{f_1} + \rho_{f_1}) + \frac{\partial p_{f_2}}{\partial \rho_{f_2}}(p_{f_2} + \rho_{f_2})\right] \tag{32g}$$

$$s = \frac{1}{(p_{f_1}+p_{f_2})}\left[\frac{\partial p_{f_1}}{\partial \rho_{f_1}}(p_{f_1} + \rho_{f_1}) + \frac{\partial p_{f_2}}{\partial \rho_{f_2}}(p_{f_2} + \rho_{f_2})\right] \tag{32h}$$

$$q = \frac{1}{2} + \frac{3}{2}\left[\frac{(p_{f_1}+p_{f_2})}{(\rho_{f_1}+\rho_{f_2})}\right] \tag{32i}$$

Here the variables can be written as follows.

$$\rho_{f_1} = 6\lambda \rho_{m0} V^{-1} \tag{33a}$$

$$\rho_{f_2} = -\frac{3B}{\alpha+3A+3} V^{\frac{\alpha}{3}} + constant * V^{-(1+A)} \tag{33b}$$

$$p_{f_1} = -2\lambda \rho_{m0} V^{-1} - 3^s \eta_0(t)(1 + 6\lambda)^s \rho_{m0}^S V^{-s} \tag{33c}$$

$$p_{f_2} = B\left[1 - \frac{3}{\alpha+3A+3}\right] V^{\frac{\alpha}{3}} + A * constant * V^{-(1+A)} \tag{33d}$$

Using the equations (32g) to (33d) we can get the variation for those parameters as follows. Fig. 35 to 40.

| 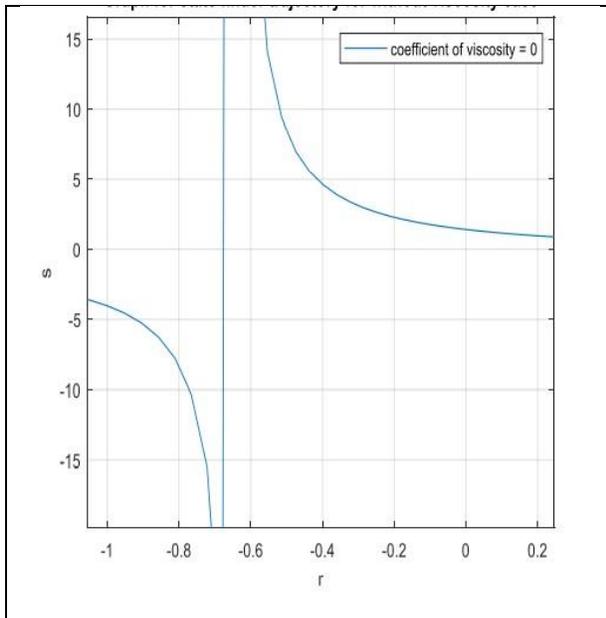 | 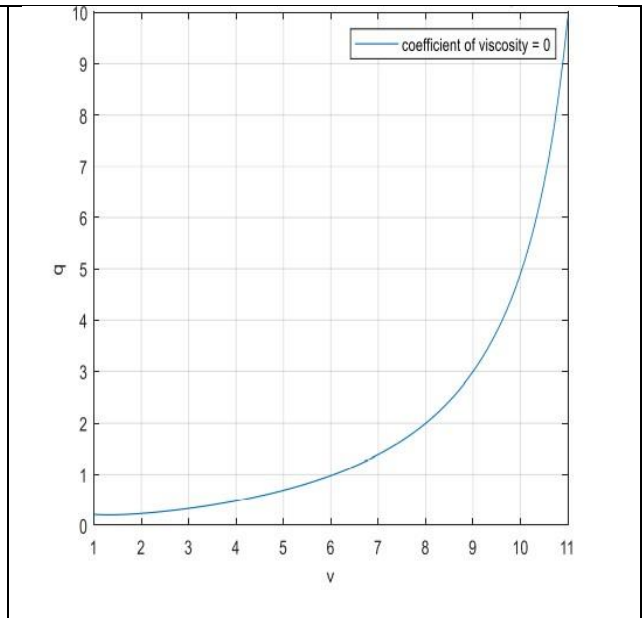 |
| --- | --- |
| Fig 35: s vs r without viscosity | Fig 36: deceleration parameter vs volume without viscosity |
| 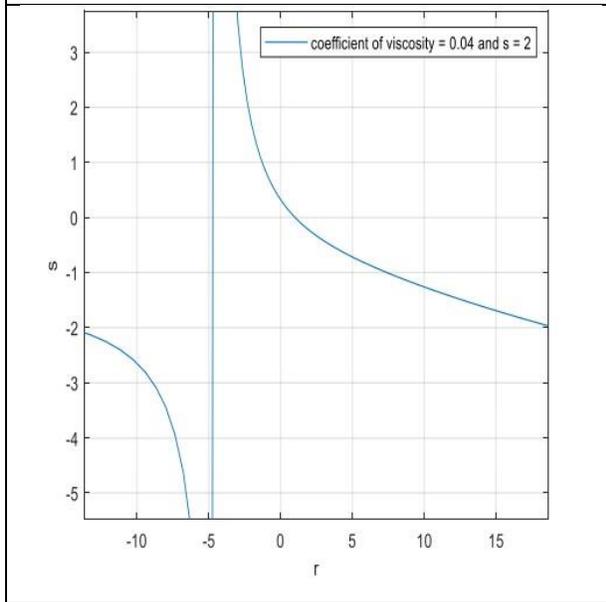 | 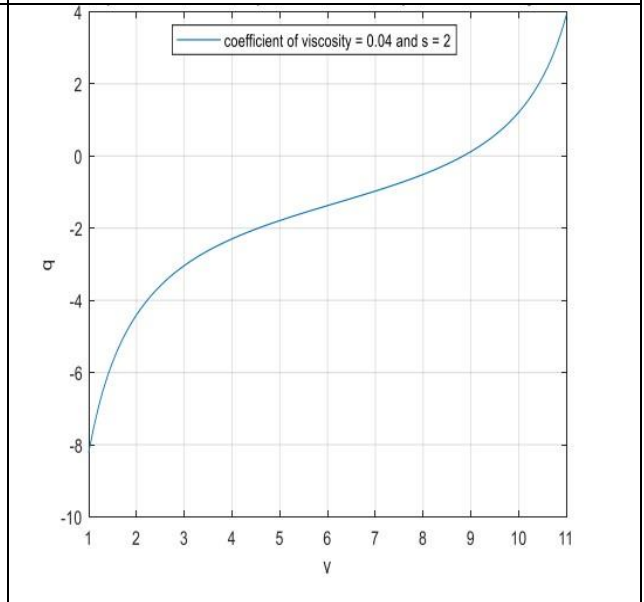 |
| Fig 37: s vs r with positive viscosity | Fig 38: deceleration parameter vs volume with positive viscosity |

| 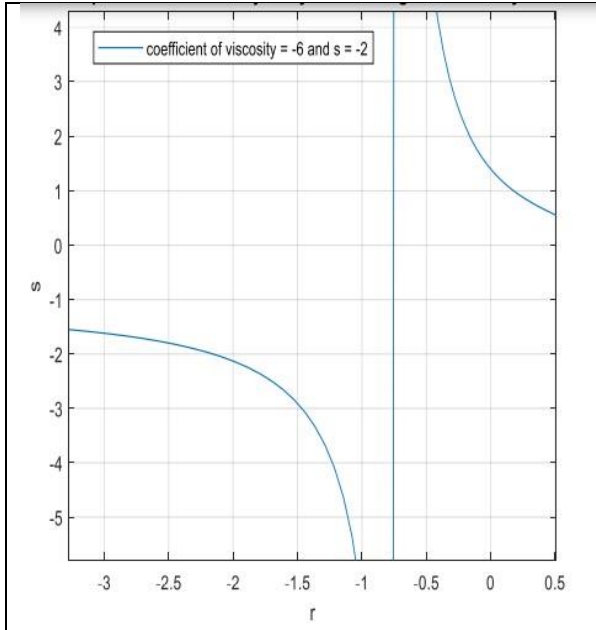 | 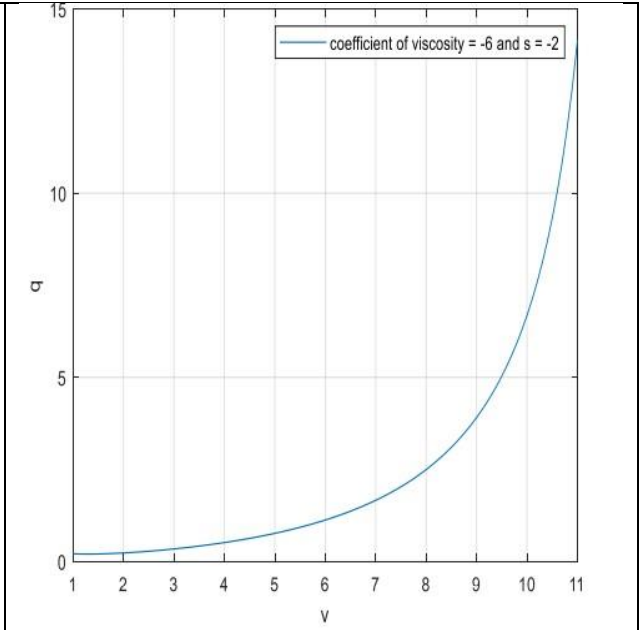 |
|---|---|
| Fig 39: s vs r for negative viscosity | Fig 40: deceleration parameter vs volume for negative viscosity |

From figures 35, 37 and 39 we can observe the nature of state finder parameter is similar. From figure 36 we observe that decelerating parameter comes out to be positive for without viscosity case i.e. it signifies decelerated expansion. For positive viscous case decelerating parameter comes out to be negative that signifies accelerated expansion (figure 38) and for negative viscous case the parameter comes out to be positive and signifies deceleration.

### 11.2. Stability analysis with cosmological perturbation:

From equations (31a) to (31c) we can proceed the calculations as follows.

$$\alpha = \frac{7}{\sqrt{3}}\sqrt{\rho_{tot}} + 2\left[\frac{\left(\frac{3}{2}-2m_1\right)\dot{\rho}_f+\left(\frac{1}{2}-2m_1\right)\dot{p}_f}{\sqrt{\dot{\rho}_f+\dot{p}_f}}\right] \tag{34a}$$

$$\mu^2 = -p_{tot} + \rho_{tot} + \frac{2}{\sqrt{3}}\sqrt{\rho_{tot}}\left[\frac{\left(\frac{3}{2}-2m_1\right)\dot{\rho}_f+\left(\frac{1}{2}-2m_1\right)\dot{p}_f}{\sqrt{\dot{\rho}_f+\dot{p}_f}}\right] \tag{34b}$$

$$c_s^2 = \frac{\frac{\partial p_f}{\partial V}}{\frac{\partial \rho_f}{\partial V}} \tag{34c}$$

From equation (34c) we may get,

$$c_s^2 = -\left[\frac{1}{3} + \frac{3^s(1+6\lambda)^s \rho_{mo}^{s-1}}{6\lambda} s\eta_0(t)V^{-(s-1)}\right] \tag{34d}$$

So, the derivations in (34b) to (34d) can be represented graphically as follows. Fig. 41 to 49.

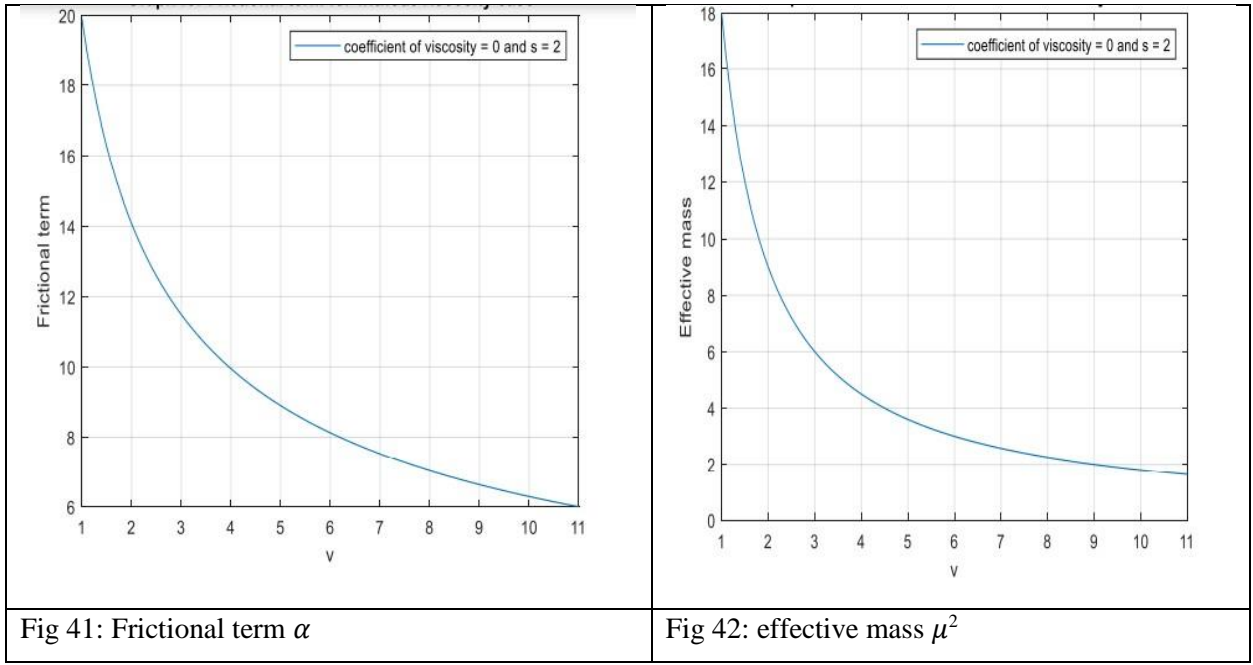

| Fig 41: Frictional term $\alpha$ | Fig 42: effective mass $\mu^2$ |

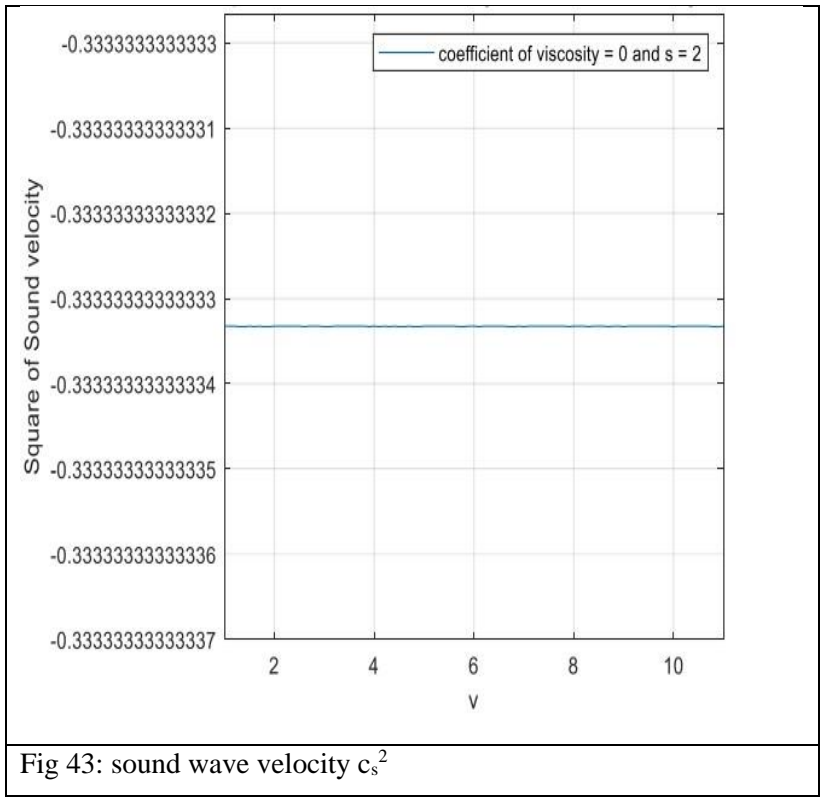

Fig 43: sound wave velocity $c_s^2$

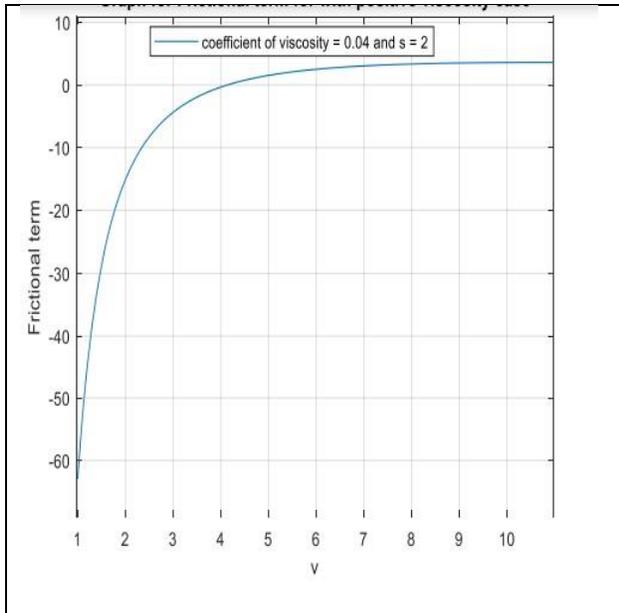

Fig 44: Frictional term with +ve viscosity $\alpha$

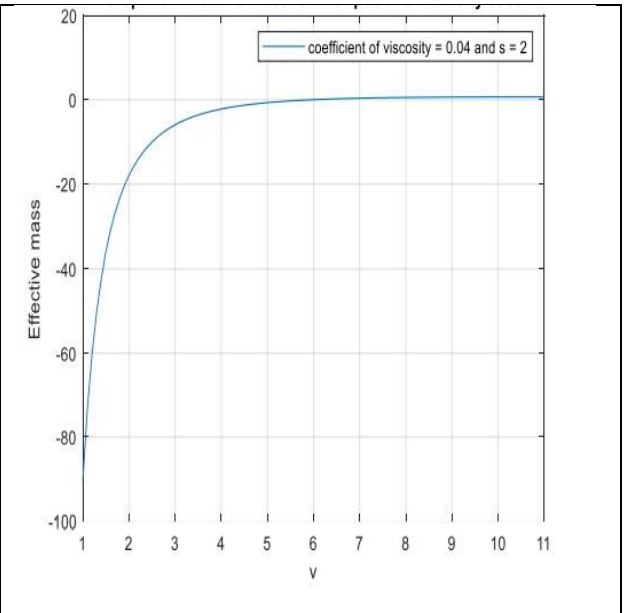

Fig 45: effective mass with +ve viscosity $\mu^2$

| 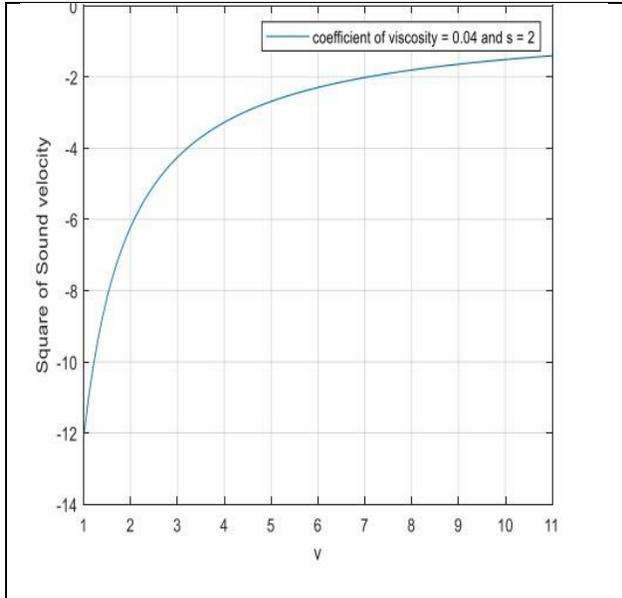 | 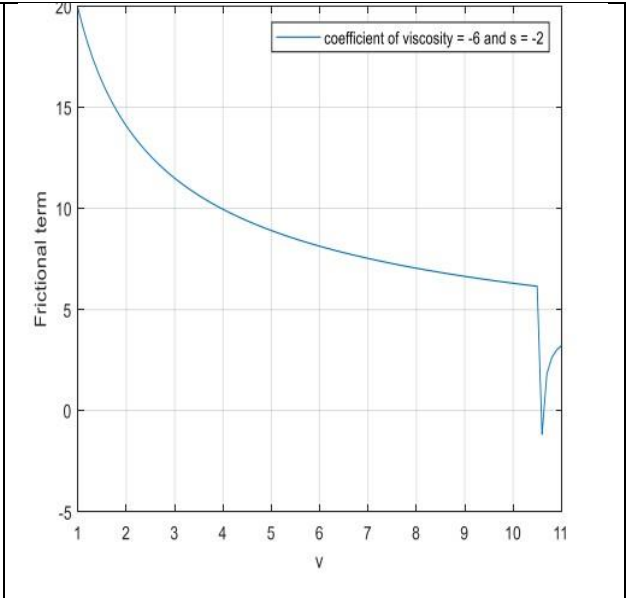 |
|---|---|
| Fig 46: sound velocity with +ve viscosity $C_s^2$ | Fig 47: frictional term with -ve viscosity $\alpha$ |
| 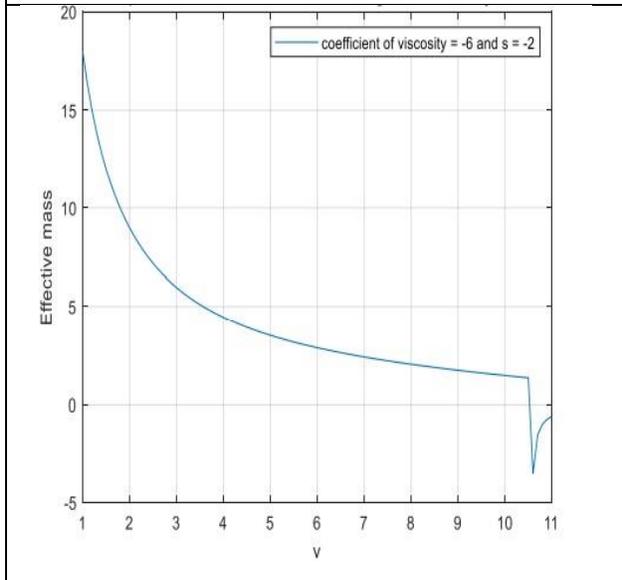 | 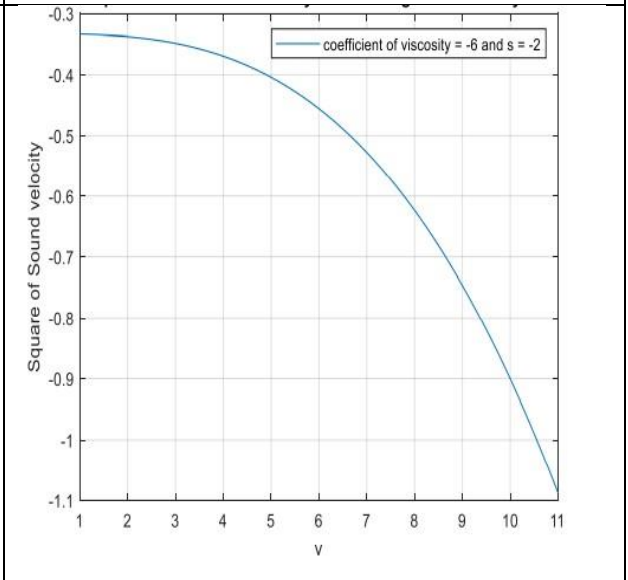 |
| Fig 48: effective mass with -ve viscosity $\mu^2$ | Fig 49: sound velocity with -ve viscosity $C_s^2$ |

## 12. Concluding remarks:

In section 7 we have introduced viscosity in order to satisfy the first two rules of thermodynamics stability of the fluid of expanding universe. Even after considering viscosity the specific heat remains negative (considering $U_0$ to be positive). In section 6 we observe that for positive viscosity the strong energy condition is violated. In order to satisfy all the energy conditions including SEC we introduced the idea of negative viscosity.

In order to resolve negative heat capacity problems, we have introduced two collateral system of fluid who are connected by an energy transition cycle. We observe that this energy transition cycle has resolved the SEC even for the positive viscosity case.

There is a resultant heat loss in cycle I in which both the systems have negative heat capacity and system 1 ($S_1$) has positive viscosity. Similarly, for cycle II the system 1 has negative heat capacity and negative viscosity but for system 2 heat capacity is positive. The resultant heat loss for each of those cycles will produce radiation pressure which pushes the SEC into positive value even the cycle I will have positive viscosity.

In section 8.3 we have discussed the possible candidates of the multiple fluid system and its thermodynamics. In section 10 we have deduced the scale factor considering the multiple fluid system.

In section 11.2 we observe that the stability parameters come out to be negative. In order to make them possible we need an additional energy $\Delta Q$ or $\Delta Q'$ which are supplied by cycle I and II. Therefore, cycle I and II are necessary to explain the evolution of universe in multiple fluid system.

**Appendix**

Here we'll discuss how we may find the deceleration or acceleration in cosmic evolution after having negative pressure including satisfaction with SEC and others conditions.

From general simplified FRW equations we get the following equation.

$$\frac{6\ddot{a}}{a} = -(\rho + 3p)$$

Here we may observe that for satisfying SEC according to Raychaudhuri equation, we must have deceleration. i.e. for $\rho + 3p \geq 0$ we must have $\ddot{a} \leq 0$.

If we introduce the cosmological constant we may write that,

$$6\frac{\ddot{a}}{a} = -(\rho + 3p) + 2\Lambda$$

Now is we may consider $2\Lambda \gg |\rho + 3p|$ (for dark energy dominated universe), then even after having satisfied SEC we can get accelerated expansion in cosmic evolution i.e. $\ddot{a} \geq 0$.

Now if $p < 0, \rho > 0$ and $|\rho| > |3p|$ then we can say inspite of having negative pressure we will have $\rho + 3p > 0$.

Hence, Raychaudhuri equation and FRW equation both have been satisfied with accelerated expanding universe with attractive gravity.

Now if we substitute cosmological constant with dark energy then we may have the following substitution.

$$2\Lambda = \rho_{DE} + 3p_{DE}$$

Hence, we may write as follows;

$$6\frac{\ddot{a}}{a} = -(\rho + 3p) + (\rho_{DE} + 3p_{DE})$$

Similarly, if $|\rho_{DE} + 3p_{DE}| \gg |\rho + 3p|$ and $\rho_{DE} + 3p_{DE} > 0$ ($p_{DE} < 0$) then Raychaudhuri equation and FRW equation both will satisfy with accelerated expanding universe with attractive gravity.

Now we'll provide another interesting phenomenon.

Suppose, we have a negative pressure of dark energy where we get $\rho_{DE} + 3p_{DE} > 0$ and $|\rho_{DE} + 3p_{DE}| \gg |\rho + 3p|$, then we will have acceleration.

Now if we'll get any phase transition that makes $\rho_{DE} + 3p_{DE} < 0$ ($p_{DE} < 0$) and $\rho_{tot} + 3p_{tot} > 0$ ($\rho_{tot} = \rho + \rho_{DE}; p_{tot} = p + p_{DE}$), then even in case of attractive gravity we get decelerated universe inspite of satisfying SEC. Although the negativity of pressure doesn't change.

So, we can see that negative pressure can provide both acceleration and deceleration in expansion.

So, we may say that negative pressure is used in physics with the only purpose to introduce the expansion in the attractive gravitation system. Thus, the expansion has been described physically. [57-72]

**Limitations of our model in this paper**

This paper consisted of investigation of a specific modified gravity model and its corresponding DBI-Essence scalar field in presence of viscous and non-viscous fluid system. We have used negative viscosity and also discussed its viability with the introduction of multi-fluid cosmology that resolved both negative viscosity and heat capacity paradox with interaction energy transition. We didn't show the nature of energy transition and resolution of present universe temperature problems is beyond the scope of this paper. The energy conditions have been discussed under the constraint that the energy density must follow the inequality as $|\rho| > |p|$. There is further scope of discussion of these limitations in our later research works. In the section of multiple fluid system, we have already introduced the idea of dark matter as a component but details of this work is beyond of the scope of this present paper where we need to introduce DM-DE interaction. This part will be given in our future works.

**Acknowledgement**

The authors are thankful to the anonymous reviewer for the supportive comments. Financial support under the CSIR Grant No. 03(1420)/18/EMRII is thankfully acknowledged by Surajit Chattopadhyay.